\documentclass[letterpaper]{emulateapj}

\usepackage{epsfig}
\usepackage{amsmath}
\usepackage{epsfig,rotating}
\usepackage{natbib}
\usepackage{lscape}
\usepackage{enumerate}
\def\sles{\lower2pt\hbox{$\buildrel {\scriptstyle <}
   \over {\scriptstyle\sim}$}}

\def\sgreat{\lower2pt\hbox{$\buildrel {\scriptstyle >}
   \over {\scriptstyle\sim}$}}
\newcommand{\cN}[1]{\mathcal{N}}

\def\gsim{\;\rlap{\lower 2.5pt
 \hbox{$\sim$}}\raise 1.5pt\hbox{$>$}\;}
\def\lsim{\;\rlap{\lower 2.5pt
   \hbox{$\sim$}}\raise 1.5pt\hbox{$<$}\;}

\slugcomment{Accepted for publication in ApJ, June 2, 2009}

\begin{document}


\title{Coupled Evolution with Tides of the Radius and Orbit of Transiting Giant Planets: General Results}

\author{Laurent Ibgui\altaffilmark{1}, Adam Burrows\altaffilmark{1}}

\affil{$^1$Department of Astrophysical Sciences, Peyton Hall, Princeton University, Princeton, NJ 08544}

\vspace{0.5\baselineskip}

\email{ibgui@astro.princeton.edu, burrows@astro.princeton.edu}

\begin{abstract}
Some transiting extrasolar giant planets have measured radii larger 
than predicted by the standard theory. In this paper, we explore the possibility that an 
earlier episode of tidal heating can explain such radius anomalies and apply
the formalism we develop to HD 209458b as an example.
We find that for strong enough tides the planet's radius can undergo a transient 
phase of inflation that temporarily interrupts canonical, monotonic shrinking due to radiative losses.
Importantly, an earlier episode of tidal heating can result in a planet with an inflated 
radius, even though its orbit has nearly circularized.  Moreover, we confirm that at late times, 
and under some circumstances, by raising tides on the star itself a planet can spiral into its host. 
We note that a 3$\times$ to 10$\times$solar planet atmospheric opacity with no tidal 
heating is sufficient to explain the observed radius of HD~209458b.
However, our model demonstrates that with an earlier phase of episodic tidal heating
we can fit the observed radius of HD 209458b even with lower (solar) atmospheric 
opacities.  This work demonstrates that, if a planet is left with an appreciable 
eccentricity after early inward migration and/or dynamical interaction, 
coupling radius and orbit evolution in a consistent fashion that includes 
tidal heating, stellar irradiation, and detailed model atmospheres
might offer a generic solution to the inflated radius puzzle for transiting 
extrasolar giant planets such as WASP-12b, TrES-4, and WASP-6b.  
\end{abstract}



\section{Introduction}
\label{sec:intro}

The most useful experimental data constraining evolutionary models of extrasolar planets 
and their radii ($R_p$) come from measurements of transiting planets\footnote{See J. Schneider's
Extrasolar Planet Encyclopaedia at http://exoplanet.eu, the Geneva Search Programme at
http://exoplanets.eu, and the Carnegie/California compilation at http://exoplanets.org.}. 
The radius of a transiting planet is inferred from transit lightcurve measurements, which 
in combination with radial velocity measurements remove the planet mass $M_{p}$ - inclination 
angle degeneracy. Much theoretical effort has been undertaken to model, and then to understand, the measured radii
\citep{guillot_et_al1996,burrows_et_al2000,Bodenheimer_et_al_2001,burrows_et_al2003,
Bodenheimer_et_al_2003,Baraffe_et_al_2003,burrows_et_al2004,Fortney_and_Hubbard_2004,
Baraffe_et_al_2004,Chabrier_et_al_2004,laughlin_et_al2005,
Baraffe_et_al_2005, Baraffe_et_al_2006,burrows_et_al2007,fortney_et_al2007,
Marley_et_al_2007,chabrier+baraffe2007,Liu_et_al_2008,Baraffe_et_al_2008}.

As \citet{burrows_et_al2007} have emphasized, custom fits are preferred for each planet
to ``project out" the effects of age, planet mass, stellar flux, etc. before one can
conclude whether a measured transit radius can or cannot be reproduced by theory.
Indeed, the transit radius of an extrasolar giant planet (EGP) depends on many parameters 
(see \citealt{burrows_et_al2007} for a sensitivity study). These include the planet mass, 
the stellar irradiation flux ${F_{p}}$, the transit radius effect \citep{burrows_et_al2003,Baraffe_et_al_2003}, 
the atmospheric composition, the presence of heavy elements in the envelope 
or in a central core, atmospheric circulation that couples the day and the 
night sides, the planet's age, and any effects that could generate an extra power 
source in the interior of the planet.  It had once been thought 
such extra power could be due to obliquity tides when the planet is in a Cassini state
\citep{winn+holman2005}, but such a possibility has now largely been ruled out 
\citep{levrard_et_al2007,fabrycky_et_al2007}. However, heating due to the 
penetration and dissipation at depth of gravity waves 
\citep{guillot+showman2002,showman+guillot2002}, or strong tidal effects 
\citep{Bodenheimer_et_al_2003,Liu_et_al_2008,Jackson_et_al_2008_1,Jackson_et_al_2008_2,Jackson_et_al_2008_3} 
have not been eliminated. The latter could be due the pumping of eccentricity by 
an undetected companion \citep{Bodenheimer_et_al_2001,Mardling_2007}.
We note that it is not known whether the tidal heat is
deposited predominantly in the convective core 
\citep{Ogilvie_and_Lin_2004,Goodman+Lackner_2008} or 
in the radiative envelope \citep{Wu_2005_1}. We make the 
default assumption of all planet radius evolutionary modelers
to date that it is deposited in the core, but the reader is 
encouraged to keep an open mind.

The initial conditions assumed for our simulations can affect the structures of EGPs at young 
ages ($\lesssim$ a few 10 Myr). As a consequence, a model of planet formation 
would be useful to determine the planet's initial entropy and, therefore, its early radius
\citep{Marley_et_al_2007,Fortney_et_al_2008}. However, since the ages of the mature transiting 
EGPs discovered so far are a few Gyrs, precise initial entropies are not crucial for our study. 
On the other hand, as we show in this paper, if tidal effects in the planet or in the star
are taken into account, the orbital parameters of eccentricity and semi-major axis left after
formation, early evolution and migration \citep{Goldreich_and_Sari_2003}, 
planet-planet scattering \citep{Ford_et_al_2003,Chatterjee_et_al_2008,Ford_and_Rasio_2008,Juric_and_Tremaine_2008}, or 
the operation of the Kozai mechanism \citep{Wu_and_Murray_2003,Wu_2003,Wu_et_al_2007,Nagasawa_et_al_2008} do matter.

There are two classes of discrepancies between measured transit radii and theoretical predictions.
The first is the subset comprised of those EGPs larger than default radius predictions and includes 
HD~209458b \citep{charbonneau_et_al2000,knutson_et_al2007a}, TrES-4 \citep{Mandushev_et_al_2007}, WASP-12b \citep{Hebb_et_al_2009},
WASP-4b \citep{Wilson_et_al_2008,Gillon_et_al_2009_1}, 
WASP-6b \citep{Gillon_et_al_2009_2}, XO-3b \citep{Johns-Krull_et_al_2008,Winn_et_al_2008},
and HAT-P-1b \citep{bakos_et_al2007a,Winn_et_al_2007,Johnson_et_al_2008}.
The other class consists of EGPs that are in fact smaller than default predictions
without dense cores and, therefore, seem to require such cores \citep{guillot06,burrows_et_al2007}.  
The correlation \citet{guillot06} and \citet{burrows_et_al2007} discovered 
between the inferred core mass and the stellar metallicity, which is not used in the modeling,
supports both the notion that dense cores are present and the core accretion mechanism for giant planet formation.

Previously, \citet{burrows_et_al2007} studied the effects of an extra 
heat source ($\dot{E}_{\rm tide}$) in the planet's interior and estimated the power 
necessary to explain some of the measured radii. \citet{Liu_et_al_2008} took this 
a step further by incorporating core tidal heating to explain the observed radii of 
TrES-4, XO-3b, and HAT-P-1b, but kept their orbital parameters constant at currently observed values. 
\citet{Jackson_et_al_2008_1,Jackson_et_al_2008_2,Jackson_et_al_2008_3} studied the coupled evolution of
$e$, $a$ and $\dot{E}_{\rm tide}$ of close-in EGPs, but maintained a constant planetary radius.

However, and importantly, \citet{Jackson_et_al_2008_1,Jackson_et_al_2008_2,Jackson_et_al_2008_3} and \citet{Gu_et_al_2003}
highlighted the possible role of tidal orbital evolution in inflating the radius of an exoplanet.
In this paper, we verify their original insight by consistently and simultaneously coupling the evolution of 
$R_{p}$, $e$, $a$, $\dot{E}_{\rm tide}$, and ${F_{p}}$ ($ \dot{E}_{\rm insolation}$) in the context
of a sophisticated atmosphere and irradiation model for the planet.
We take into account the tides raised on the planet and the tides raised on the star. 
We find that if the tides are strong enough a planet's radius 
can undergo a transient phase of inflation that temporarily interrupts shrinking, 
and does so at epochs consistent with the few Gyr ages of transiting planets.
Extremely strong tides fade at a very early stage and have a negligible impact 
on the radius at Gyr ages.  It is noteworthy that the behavior is nonlinear 
and depends sensitively on the initial orbital conditions and on tidal heating parameters.
In addition, we find that increasing the planet's atmospheric opacity accelerates and increases
the magnitude of the transient effect. If the tides raised on the star are negligible, 
the orbit reaches a final circular equilibrium state.  Otherwise, there is no final 
equilibrium, and the planet might eventually plunge into its host star \citep{rasio_1996,Levrard_et_al_2009}.
We find that due to an earlier phase of tidal heating a planet whose orbit has 
circularized can still have an inflated radius.  As an example and proof of principle, we 
apply our formalism to HD~209458b. We find a set of tidal parameters, $Q'_{p}$ and $Q'_{\ast}$, and
initial orbital parameters, $e_{i}$ and $a_{i}$, that lead to the measured values of $R_{p}$, $a$, $e$, even
for atmospheric opacities for solar equilibrium abundances. 
We find that HD~209458b's radius can also be fit with no tides and a 3$\times$ to 10$\times$solar 
planet atmospheric opacity at the new and shorter age of $\sim$3.1 Gyrs suggested by \citet{Torres_et_al_2008}.
However, unless the atmospheric abundances can be measured, it will be
difficult to distinguish the different predictions, particularly given the
current ambiguity in $Q'$.  Nevertheless, if there are independent (theoretical?)
constraints on $Q'$, and/or precision measurements of the atmospheric spectra,
one might be able to discriminate between the two different explanations.

In \S\ref{sec:model}, we present our formalism, model assumptions, and computational techniques.
Section \ref{sec:generic} is the central section of the paper in which we describe 
the generic coupled evolution of $R_{p}$, $e$, $a$, $\dot{E}_{\rm tide}/ \dot{E}_{\rm insolation}$ 
when tides play a significant role. We describe in \S\ref{subsec:typical_case} the typical 
scenario, which exhibits transient radius inflation due to tidal heating of the planet. 
The case of extremely strong tides is discussed in \S\ref{subsec:strong_tides}.
In \S\ref{subsec:atm_opacity}, we address the effect of atmospheric opacity and in 
\S\ref{subsec:tides_star} we analyze the influence of the tides raised on the star. 
The latter can cause the planet to spiral in. In \S\ref{sec:HD209458b}, we apply the model 
to HD~209458b.  Finally, in \S\ref{sec:conclusion} we summarize our results, review caveats,
and discuss how such transient tidal heating might explain other large-radius EGPs, such as WASP-12b, TrES-4, and WASP-6b.

\section{Model: Formalism, Assumptions, and Computational Techniques}
\label{sec:model}

The evolution of a planet's eccentricity and semi-major axis due to tidal effects
depends sensitively on it radius.  In turn, the evolution of its radius
depends sensitively on the tidal heating power in its interior associated with 
the corresponding evolution of its orbital parameters.  Moreover, a planet's radius 
evolution is sensitive to the degree of stellar irradiation, which is directly tied
to the planet-star distance.  Therefore, to perform simultaneous orbital and radius
evolutions, all the relevant equations must be coupled and atmospheric boundary conditions
that vary systematically with changing irradiation regimes must be incorporated. 
We have established the tools necessary to self-consistently accomplish such
calculations and in this section we describe our methods.

We assume that the planet has a spherical gaseous $\rm H_{2},He$ envelope, and 
use the equation of state of \citet{Saumon_et_al_1995}. The helium mass fraction 
($Y$) is set equal to $0.25$. The effect of reasonable variations in $Y$ on the radius of an EGP is small.
Except in the atmosphere, we assume the planet is fully convective and that its envelope contains no heavy elements.
Though our formalism allows it, in this paper we ignore the posssible effects of an inner dense core.  
As shown, for instance, in \citet{guillot06} and \citet{burrows_et_al2007}, the presence of heavy elements in 
either the envelope or the core decreases the total radius of the EGP.
To model the evolution of the planet, we use the Henyey evolutionary code of \citet{Burrows_et_al_1993,Burrows_et_al_1997},
with boundary conditions that incorporate realistic irradiated planetary atmospheres. Using COOLTLUSTY, a variant of the spectral
atmosphere code TLUSTY \citep{Hubeny_and_Lanz_1995}, we precalculate grids of atmospheres for various values of $T_{eff}$ 
and surface gravity $g$.  These provide the associated entropy $S$ in the convective region 
of the planet. Inverting the relation to obtain $T_{eff}(S,g)$ yields the energy flux escaping the interior
(``$\sigma T_{eff}^4$"; \citealt{burrows_et_al2003}). We precalculate these grids not 
only for a given gravity and $T_{eff}$, but also for each orbital pair $(e,a)$. The stellar flux 
at the planet is taken to be the time-averaged mean during an orbit.  Thus, we have atmospheric 
and evolutionary boundary conditions which follow the evolution of the planetary orbit.  
The stellar spectrum is interpolated at the actual effective temperature and gravity of the 
star from the Kurucz stellar atmosphere models \citep{Kurucz_1994}. Note that we 
assume the stellar spectrum is constant during evolution. We include the 
``transit radius effect," which accounts for the fact that the transit 
radius is an impact parameter \citep{burrows_et_al2003,Baraffe_et_al_2003}.

In our calculations, we assume that tidal heating occurs entirely in the convective interior 
of the planet and that the evolutionary process starts a few Myr after the star's formation. 
Therefore, we suppose that the protoplanetary disk has dissipated 
\citep{Goldreich_and_Sari_2003} and that any ensuing chaotic collisional period of planet-planet scattering
has ended \citep{Ford_et_al_2003,Juric_and_Tremaine_2008,Chatterjee_et_al_2008,Ford_and_Rasio_2008,Nagasawa_et_al_2008}.
Importantly, we presume that, after these early formation, migration, and dynamical 
phases, an interesting subset of close-in EGPs are left with high values of $e_{i}$ ($\gtrsim0.2$)
and small values of the ``initial" semi-major axes, $a_i$ ($\sles 0.1-0.15$~AU). ($a_i$
is still larger than the ``final" values currently observed.)  These assumptions
are necessary for tidal effects to be of interest as a possible explanation
for the large planetary radii observed in a subset of cases and 
are not unreasonable \citep{Ford_and_Rasio_2008,Nagasawa_et_al_2008}.   
We neglect stellar and planetary obliquities, assume that the planet's spin is 
synchronized (is tidally locked) with its orbital period, and that the star's spin rate
is small compared with the orbital mean motion. A rough estimate of the synchronization 
time ($\tau_{sync}$) for a close-in EGP gives a value between one and a few$\times$10 Myr \citep{guillot_et_al1996}. 
We assume that equilibrium tides have a constant lag angle for 
any frequency and that $Q'$ is independent of orbital period.
{Another approach is to assume a constant time lag for any 
frequency, as was done in the pioneering work of \citet{Darwin_1880}
(see also \citealt{Hut_1981}, \citealt{rasio_1996}, \citealt{Eggleton_et_al_1998}, and \citealt{Levrard_et_al_2009}). 
Higher-order $e$ formulations, though developed 
(\citealt{Mardling_and_Lin_2002,Ogilvie_and_Lin_2004,dobbs-dixon+lin2004,Mardling_2007}), 
introduce numerous other uncertain approximations concerning tidal processes.
Given our current limited knowledge of tidal dissipation/heating in EGPs, 
we feel the approach we have employed is acceptable.} 

With all these assumptions, the equations of tidal evolution of eccentricity $e$, semi-major axis $a$,
and tidal heating rate $\dot{E}_{\rm tide}$, to second order in eccentricity, are
\citep{Goldreich+Soter_1966,Kaula_1968,Peale_and_Cassen_1978,
Murray_et_Dermott_1999,Bodenheimer_et_al_2001,Bodenheimer_et_al_2003,Gu_et_al_2004,Mardling_2007,
Jackson_et_al_2008_2,Jackson_et_al_2008_3,Jackson_et_al_2008_4,Ferraz-Mello_et_al_2008,
Barnes_et_al_2009}:
\begin{alignat}{2}
\frac{1}{e} \frac{de}{dt} &= - \frac{1}{a^{13/2}}  \left[ \overbrace{ \phantom{2} K_{1p_{\ast}} \frac{R_{p}^{5}}{Q'_{p}} \phantom{~e^{2}} } {}+
                                             \overbrace{ \phantom{ \frac{8}{25} \left( 1+\frac{57}{4}~e^{2} \right) {}} K_{2p_{\ast}} \frac{R_{\ast}^{5}}{Q'_{\ast}} } \right], \label{eq:e} \\
\frac{1}{a} \frac{da}{dt} &= - \frac{1}{a^{13/2}}  \left[\underbrace{  2 K_{1p_{\ast}}    \frac{R_{p}^{5}}{Q'_{p}}~e^{2} }_{\rm{tides~on~planet}} {}+
                                                      \underbrace{ \frac{8}{25} \left( 1+\frac{57}{4}~e^{2} \right) K_{2p_{\ast}} \frac{R_{\ast}^{5}}{Q'_{\ast}} }_{\rm{tides~on~star}}  \right]\, , \label{eq:a}
\end{alignat}

and

\begin{eqnarray}
\dot{E}_{\rm tide}           &= & \phantom{-}  \left( \frac{63}{4} G^{3/2}M_{\ast}^{5/2} \right)  \frac{R_{p}^{5}}{Q'_{p}}  \frac{e^{2}}{a^{15/2}}\, ,                   \label{eq:Etides} 
\end{eqnarray}

\noindent where $K_{1p_{\ast}}$ and $K_{2p_{\ast}}$ are constants defined by
\begin{eqnarray}
K_{1p_{\ast}} &= & \frac{63}{4} G^{1/2} \frac{M_{\ast}^{3/2}}{M_{p}} \\
K_{2p_{\ast}} &= & \frac{225}{16} G^{1/2} \frac{M_{p}}{M_{\ast}^{1/2}}\, ,
\label{eq:Kp_Ks}
\end{eqnarray}

\noindent $G$ is the gravitational constant, and $M_{p},~M_{\ast},R_{p},~R_{\ast}$ 
are the masses and radii of the planet and star. Note that the planet radius is 
time-dependent ($R_{p}\left( t \right)$), but the star's radius is assumed to be constant.
The evolution of $e$ and $a$ are due both to tides raised on the planet (due to the star) 
and tides raised on the star (due to the planet). Equations (\ref{eq:e}) and (\ref{eq:a}) 
indicate that $e$ and $a$ can only decrease. This is because we assume the stellar spin rate is low.
This is different from, for example, the Earth-Moon system, for which the tides raised 
on the Earth increase the semi-major axis, while the tides raised 
on the Moon decrease it \citep{Goldreich+Soter_1966}.  As \citet{Mardling_2007} has pointed out,
eqs. (\ref{eq:e}), (\ref{eq:a}), and (\ref{eq:Etides}) don't include the higher-order terms
beyond $e^2$ and these terms should be important for high $e$.  However, 
in this paper we focus on the generic character of the planet-orbit coupling 
and its potential role in explaining the subset of EGPs with anomalous 
radii.  In this light, and given the many remaining ambiguities in  
tidal heating theory, our formalism should be adequate to address 
the central phenomena.

The $Q'$s in eqs. (\ref{eq:e}), (\ref{eq:a}), and (\ref{eq:Etides}) are the tidal 
dissipation factors ($Q'_{p}$ in the planet and $Q'_{\ast}$ in the star), 
given by $Q'=3Q/2k_{2}$, where $Q$ is the specific tidal dissipation 
function \citep{Goldreich_1963} and $k_{2}$ is the Love number 
\citep{Love_1927,Goldreich+Soter_1966,Ogilvie_and_Lin_2007}.
Smaller $Q'$s result in higher rates of tidal dissipation.
For Jupiter, we have $Q' \approx 4~Q$ \citep{Gavrilov_and_Zharkov_1977}
and estimates of $Q$ are $\sim$$6\times10^{4}-10^{6}$ \citep{Goldreich+Soter_1966,Yoder_and_Peale_1981}. 
\citet{Ogilvie_and_Lin_2004} have provided a theoretical motivation
for values of $Q$ near $\sim 10^{5}$.  For a synchronized short-period EGP, we
have only crude estimates, but \citet{Ogilvie_and_Lin_2004} propose values of $\sim 5\times10^{6} - 10^{7}$
and \citet{Jackson_et_al_2008_1,Jackson_et_al_2008_2} propose values of 
$Q'_{\ast} \sim 10^{5.5}$ and $Q'_{p} \sim 10^{6.5}$. The latter are based on the statistical
distribution of initial values of $e$ and $a$ the authors obtain after integrating eqs. 
(\ref{eq:e}) and (\ref{eq:a}) backwards in time at constant $R_{p}$.
For terrestrial planets, the physics of tidal dissipation is different and
it is thought that $Q' \gtrsim 10^{2}$ \citep{Dickey_et_al_1994,Mardling_and_Lin_2004,Jackson_et_al_2008_4}.
Most of the numerical values for this factor are empirically determined.
For Jovian planets and EGPs there are various ongoing efforts to improve our understanding
of tidal dissipation \citep{Ogilvie_and_Lin_2004,Wu_2005_1,Wu_2005_2,Goodman+Lackner_2008},
but there is as yet no comprehensive theory.  Moreover, theory has yet to 
determine where in the planet the tidal heat is deposited.  If it is not deposited 
in the convective core, but in the radiative atmosphere, our results 
might need to be altered.  \citet{Goodman+Lackner_2008} 
have studied dynamical tides in an isentropic fluid body, focusing on 
inertial waves excited by scattering from the rigid core of a planet,
and conclude that these waves might dissipate most of their energy 
in the convective region.  This is also the suggestion of \citet{Ogilvie_and_Lin_2004}.
\citet{Wu_2005_1}, on the other hand, suggests that a large fraction 
of the tidal heat could be deposited in the outer regions, perhaps 
in the radiative zone.  However, all modelers who have explored the possible
effects of tidal heating on EGP radius evolution have assumed that the heat is deposited
in the convective interior where it can be redistributed almost instantly.
For specificity, we stick to this convention for this study, 
assume that $Q'$ is a property of each body, and assume it 
is constant during the integration.

Since radius evolution must account for stellar irradiation, 
an informative quantity is the ratio between the tidal heating rate $\dot{E}_{\rm tide}$ and 
the insolation rate, $ \dot{E}_{\rm insolation}= \pi R_{p}^{2}~F_{p}$, 
where $F_{p}$ is the flux at the substellar point. To second order in the 
eccentricity, this ratio is:
\begin{eqnarray}
\frac{\dot{E}_{\rm tide}}{ \dot{E}_{\rm insolation}} &= & \left( \frac{63}{4\pi\sigma}~G^{3/2} \right) \left( \frac{M_{\ast}^{5/2}}{R_{\ast}^2~T_{\ast}^{4}} \right) 
                                                  ~\frac{R_{p}^{3}}{Q'_{p}}~\frac{e^{2}}{a^{11/2}} \, ,
\label{eq:Etides_Einsol}
\end{eqnarray}
where $\sigma$ is the Stefan-Boltzmann constant and $T_{\ast}$ is 
the effective temperature of the star. As is clear from eq. (\ref{eq:Etides_Einsol}),
this ratio is a very stiff function of $a$, and a moderately stiff function of $R_{p}$. 
We integrate forward in time equations (\ref{eq:e}) and (\ref{eq:a}), along with the Henyey 
equations of planetary structure and radius evolution which incorporate 
tidal heating given by eq. (\ref{eq:Etides}).  We start 
at various arbitrarily specified initial eccentricities $e_{i}$ and 
semi-major axes $a_{i}$ and also specify $Q'_{p}$ and $Q'_{\ast}$, keeping 
them constant during an integration. The strong nonlinear coupling between ${e}$ and ${a}$, 
already emphasized by \citet{Jackson_et_al_2008_2}, is made all the more 
so in our more general formalism by the variation in $R_p$ and the stiff dependence
on $R_p$ in eqs. (\ref{eq:e}) and (\ref{eq:a}).  

From eqs. (\ref{eq:e}) and (\ref{eq:a}), we can estimate the relative 
contributions of the tides raised on the planet (due to the star)
[$\left( de /dt \right) _{P}$, $\left( da /dt \right) _{P}$] 
and raised on the star (due to the planet) 
[$\left( de /dt \right) _{\ast}$, $\left( da /dt \right)_{\ast}$]. 
We obtain: 
\begin{alignat}{2}
\frac {\left( de /dt \right) _{P}}{\left( de /dt \right) _{\ast}}  &={} &\frac{28}{25}
           &\left(\frac{M_{\ast}}{M_{P}} \right)^{2} \left(\frac{R_{P}}{R_{\ast}} \right)^{5} \left(\frac{Q'_{\ast}}{Q'_{P}} \right), \label{eq:eS_eP} \\
\frac {\left( da /dt \right) _{P}}{\left( da /dt \right) _{\ast}}  &={} & \left( \frac{7e^{2}}{1+\frac{57}{4}e^{2}} \right)  &\left(\frac{M_{\ast}}{M_{P}} \right)^{2} 
\left(\frac{R_{P}}{R_{\ast}} \right)^{5}  \left(\frac{Q'_{\ast}}{Q'_{P}} \right)  \label{eq:aS_aP}\, .
\end{alignat}

\noindent The factor $\left[7e^{2} / (1+57e^{2}/4) \right]$ in eq. (\ref{eq:aS_aP}) 
ranges from $\simeq 0.46$ ($e=1$) to zero ($e=0$). If we assume that the EGP has 
the mass and the radius of Jupiter and orbits a sun-like host star, then the 
ratio given by eq. (\ref{eq:eS_eP}) is  $\left( \dot{e}_{P} / \dot{e}_{\ast} \right) 
\simeq 14 \left( Q'_{\ast}/ {Q'_{P}} \right)$ and the ratio given by eq. (\ref{eq:aS_aP}) is
$\left( \dot{a}_{P} / \dot{a}_{\ast} \right) \simeq 88 \left( Q'_{\ast}/ {Q'_{P}} \right) \left[ e^{2} / (1+57e^{2}/4) \right] $.
For HD~209458b, the factors 14 and 88 become 73 and 457, respectively. Thus, if the 
star is as dissipative as the planet ($Q'_{\ast} \sim Q'_{P}$) or less dissipative 
($Q'_{\ast} > Q'_{P}$), then the evolution of the eccentricity is due mainly to 
the effect of the tides raised on the planet. However, if the dissipation inside the 
star is high enough ($Q'_{\ast} \ll Q'_{p}$), then the effect of the tides 
raised on the star has a comparable or even greater influence.  Moreover, once the 
orbit is circular ($e$ = 0), the tidal effect on the planet ceases (eq. \ref{eq:Etides}),
but since the planet and star are not synchronized the tidal effect on the star does not. 
As a consequence, the planet's orbital decay continues, though on a different timescale
and it can eventually inspiral into its host star.  Incidentally, when $e=0$, eq. (\ref{eq:a}) 
can be integrated analytically to yield:
\begin{eqnarray}
a &= & a_{0}\left[ 1 - \frac{117}{4} \frac{G^{1/2}}{a_{0}^{13/2}} \frac{M_{P}}{M_{\ast}^{1/2}} ~ \frac{R_{\ast}^5}{Q'_{\ast}} \left(t-t_{0} \right)  \right]^{2/13} ,
\label{eq:a_when_e_equal_0}
\end{eqnarray}
\noindent where $a_{0}$ is the semi-major axis at any time $t_{0}$ after the orbit 
has circularized. A similar formula is given in \citet{Goldreich_1963}. This formula demonstrates
that spiral in accelerates when the star is more dissipative ($Q'_{\ast}$ lower),
has a bigger radius $R_{\ast}$, has a lower mass $M_{\ast}$, or if the planet has a higher mass $M_{p}$.

\section{The Generic Coupled Evolution of the planetary radius, the eccentricity, and the semi-major axis}
\label{sec:generic}

In this section, we present and analyze the generic results of interest that have 
emerged from our calculations of the simultaneous evolution of the radius and orbital 
parameters of transiting (close-in) EGPs when tidal effects are included. The models 
assume various values of $Q'_{p}$ and $Q'_{\ast}$, as well as representative initial 
values of the orbital parameters $e_{i}$ and $a_{i}$, and are chosen to highlight various 
possible behaviors that might be germane to the explanation of the anomalous radii
of a subset of the measured transiting EGPs. Moreover, for discussions 
of a ``generic transiting system" we have employed the properties of 
HD~209458 and HD~209458b (listed in Table \ref{tab:HD209458_data}).
However, it is important to point out that the values of the relevant
parameters are specific to each individual planet-host star system.  
In each of the following subsections, we illustrate our conclusions 
with a four-panel figure (Figs. \ref{fig:ms_evolution_fig1}, \ref{fig:ms_evolution_fig2},
\ref{fig:ms_evolution_fig3}, and \ref{fig:ms_evolution_fig4}) 
in two rows, with $R_{p}(t)$ and $e(t)$ on the first row, 
and $a(t)$ and $(\dot{E}_{\rm tide} /  \dot{E}_{\rm insolation})[t]$
on the second row.

\subsection{A Baseline Scenario}
\label{subsec:typical_case}

For clarity's sake, we here neglect the influence of tides raised 
on the star (see \S\ref{subsec:tides_star}), equivalent to setting $Q'_{\ast} \rightarrow \infty$. 
Figure~\ref{fig:ms_evolution_fig1} depicts the simultaneous evolution of the 
planet's radius $R_{p}(t)$, eccentricity $e(t)$, semi-major axis $a(t)$, 
and power ratio $(\dot{E}_{\rm tide} /  \dot{E}_{\rm insolation})[t]$ and
demonstrates how tides raised on a close-in planet might induce a transient 
phase of radius inflation, with consequences on Gyr timescales.
In this baseline case, $Q'_{p}=10^{6.5}$ and $a_{i}=0.075$~AU. We plot 
curves for different initial eccentricities ($e_{i}=0.65,0.60,0.55,0.40,0.30,0.20$)
and in Fig. \ref{fig:ms_evolution_fig1} use atmospheric opacities for solar-metallicity equilibrium 
chemical compositions. For comparison, we include two curves depicting cases with no tides 
and, therefore, no tidally-induced migration, zero eccentricity, but two different initial/final 
values of the semi-major axis (black dotted: $a=a_{i}=0.075$~AU;
black dashed: $a=0.047$~AU, the current position of HD 209458b). 

Typical behavior is illustrated by the curves on Fig. \ref{fig:ms_evolution_fig1}, for 
which $e_{i}=0.65,0.60,0.55$.  The very first episode of radius shrinkage is 
followed by a transient period of expansion, after which the radius resumes shrinking at 
a progressively slower rate.  Let us describe the case for which $e_{i}=0.65$.
When $e_{i}=0.65$, radius expansion starts at the age of $\sim$0.13 Gyr 
($\rm R_{p} \simeq 1.48~R_{J}$), almost in phase with the increase 
in the power ratio ($\rm \simeq 3 \times 10^{-3}$).
The peak of the radius is reached at $\sim$0.85~Gyr ($\rm R_{p} \simeq 1.78~R_{J}$), 
and roughly coincides with the peak of the power ratio ($\rm \simeq 8 \times 10^{-3}$), 
reached at $\sim$0.80~Gyr. The corresponding time lag of $\sim$50~Myr is not universal, but is generally 
shorter than the characteristic thermal cooling timescale.  The latter depends 
more directly on diffusion through the thick atmosphere, while the former is more
dependent on core tidal heating, to which the radius responds more directly.     
The average rate of radius expansion from the starting point of this phase 
to the peak is roughly $\rm 0.43 ~ R_{J} ~ Gyr^{-1}$. The corresponding average 
rate of increase in the power ratio is $\rm \sim 7 ~ Gyr^{-1}$. After the peak in the 
power ratio is achieved, it is followed by a drop at a rate of $\sim$$4 \times 10^{-3}~{\rm Gyr}^{-1}$. 
The rate at which $R_{p}$ decreases is comparable to the rate at which it increases,
until an age of $\sim$1.2~Gyr, after which it progressively flattens and tends to zero, 
with an average value between 5 and 6~Gyr of $\sim$$\rm 0.01 ~ R_{J} ~ {\rm Gyr}^{-1}$.
It is important to reemphasize that all the variables 
$R_{p}$, $e$, $a$, and $\dot{E}_{\rm tide} /  \dot{E}_{\rm insolation}$ are interdependent
and evolve consistently by mutual influence. Thus, as suggested by eq. (\ref{eq:Etides_Einsol}),
the peak in the power ratio is the result of the combined evolution of ${e}$, ${a}$, and ${R_{p}}$, 
which in turn depends on this ratio. During the radius inflation phase, 
$e$ and $a$ decrease at a rate that increases in absolute value.
The peak value of the planet's radius corresponds closely to an inflection point in the evolution 
of $e$ and $a$.  After this inflection point, the absolute value 
of the rates of change of both $e$ and $a$ decrease, finally 
tending to zero.  

While the circularizing orbit is gradually moving closer to the star, 
the stellar irradiation flux ${F_{p}}$ is increasing and its role in 
stanching heat loss from the planet and slowing radius shrinkage is strengthened \citep{burrows_et_al2000}. 
Once orbital equilibrium state has been achieved, tidal effects disappear
and $R_p$ continues to evolve due to radiative losses from the surface \citep{burrows_et_al2007}.
Therefore, due to an earlier episode of tidal heating a planet's orbit can currently be 
circular, and yet its radius can be larger than the default evolutionary theory would predict. 
In other words, a zero eccentricity orbit and, therefore, the absence 
of current tides, does not preclude an inflated radius due to the earlier 
action of tides.  We will see in \S\ref{subsec:tides_star} that this equilibrium orbital state 
might not persist if we incorporate tides raised on the star and the corresponding 
$Q'_{\ast}$ is small enough. Note that even if the maximum ratio $\dot{E}_{\rm tide}/ \dot{E}_{\rm insolation}$ 
can appear quite small (e.g., $\sim 10^{-2}$), it can have a large effect 
on the orbital parameters and radius evolution \citep{Liu_et_al_2008}. 
%
%

For lower values of $e_{i}$, the duration of the transient phase increases, 
the maximum $R_p$ achieved decreases, and the epoch of peak $R_p$ shifts to older ages.
The peak disappears altogether for values of $e_{i}$ below between $0.40$ and $0.30$. 
However, this does not mean that tidal effects are no longer important. 
For lower values of $e_{i}$, tidal effects still slow the decrease in $R_p$.
All else being equal, in particular for a given $a_{i}$, the lower the initial 
eccentricity $e_{i}$, the slower the evolution of the eccentricity to zero and the slower the evolution of
the semi-major axis. In other words, the lower the value of $e_{i}$, the slower the evolution
of both $e$ and $a$ and the higher the final $a$, for a given $a_i$.  As for its 
impact on radius evolution, a lower $e_{i}$ implies a weaker, but longer lasting,
tidal influence. If $e_{i}$ is small enough, the planet no longer receives 
enough tidal heat to inflate.  All these outcomes are straightforwardly understood 
as consequences of eqs. (\ref{eq:e}), (\ref{eq:a}), (\ref{eq:Etides}), and (\ref{eq:Etides_Einsol}), 
with $Q'_{\ast} \rightarrow \infty$.  Thus, a lower value of $e_{i}$ implies lower values of 
$\dot{E}_{\rm tide}$ and, therefore, less power to inflate the planet   
or reduce its rate of shrinkage. Clearly, it also implies a lower initial power ratio 
$\dot{E}_{\rm tide} /  \dot{E}_{\rm insolation}$ and a slower rate of initial 
decrease of $e$ and $a$.  Furthermore, numerical integration of the set of evolutionary 
equations shows that this initial trend for $e$ and $a$ persists at later ages, 
even if the radii and power ratios ``invert."  By this we mean that the radius 
at around 1 Gyr when ${e_{i}=0.65}$, for example, can be much larger than the radius when ${e_{i}=0.20}$. 
However, after $\sim$3.5 Gyr it then decreases much faster and ends up smaller. 
Nevertheless, there is a small, but long lasting, tidal heating effect even 
in the case of $e_{i}=0.20$. Eventually, the final radii are larger if the 
final semi-major axes are smaller, consistent with the conclusions of \citet{burrows_et_al2007} $-$ 
for a circular orbit, the closer the planet is to the star, the higher 
the insolation flux, and the larger its radius at a given epoch.

Comparing the case without tides, but with $a=a_{i}=0.075$~AU (black dotted curve 
on Fig. \ref{fig:ms_evolution_fig1}), with those with non-zero values of $e_{i}$ and 
the same $a_i$, demonstrates that the planet's radius would always be larger when the planet 
can be tidally heated. However, even without tides the radius of a planet 
already at its current position (e.g., $a=0.047$~AU, 
for HD 209458b) can be larger than the radius of a planet 
which experiences tidal heating. Fig. \ref{fig:ms_evolution_fig1} demonstrates that,
if the planet starts at a larger orbital distance, tidal heating does 
not necessarily and universally result in a larger planetary 
radius at a given age. 
%

\subsection{Extremely Strong Tidal Effects}
\label{subsec:strong_tides}

As described in \S\ref{subsec:typical_case}, tidal heating can have 
a significant, at times non-monotonic, impact on a close-in EGP's radius evolution.
This is a key conclusion.  However, contrary to what one might naively expect, extremely 
strong tidal dissipation in a planet can have a negligible effect on its late-time radius.
Figure~\ref{fig:ms_evolution_fig2} depicts the simultaneous evolution of 
a planet's radius $R_{p}(t)$, eccentricity $e(t)$, semi-major axis $a(t)$, 
and power ratio $(\dot{E}_{\rm tide} /  \dot{E}_{\rm insolation})[t]$ under such circumstances.
For comparison, the same two tide-free cases depicted in Fig. \ref{fig:ms_evolution_fig1} 
are also plotted in Fig. \ref{fig:ms_evolution_fig2}. 
The only difference with what was discussed in \S\ref{subsec:typical_case} is that we assume
$Q'_{p}=10^{5.0}$, instead of $Q'_{p}=10^{6.5}$.  We still assume $a_{i}=0.075$~AU 
and ignore stellar tides. Models with only three of the initial eccentricities 
from Figure~\ref{fig:ms_evolution_fig1} are shown ($e_{i}=0.40,0.30,0.20$), sufficient
to demonstrate our point.  Note that the plots focus on an earlier interval (2 Gyr).

In this case, after only $\sim$10 Myr, whatever the initial eccentricity, $R_p$  
inexorably shrinks and the associated curves are much closer.  This is despite 
the fact that in the early stages of evolution the power ratio, for 
$e_{i}=0.40$ for example, is two orders of magnitudes higher than in the 
baseline scenario of \S\ref{subsec:typical_case} $-$ $10^{-1}$ compared with $10^{-3}$.
However, Fig. \ref{fig:ms_evolution_fig2} indicates that such large additional core power
has had only a small effect on the radius of the planet at the typical ages of observed transiting EGPs, i.e. at a few Gyr.
This is because the orbit circularizes very rapidly and the tidal power ratio disappears extremely quickly.
The final semi-major axis is reached in less than $\sim$0.1 Gyr and the eccentricity is damped in less than $\sim$0.8 Gyr.
If we look at ages earlier than $\sim$1 Gyr (not shown), we do see a transient phase of radius inflation for the case $e_{i}=0.40$.
The peak occurs very early ($\sim$10 Myr), is of very short duration, and though 
the planet's radius can reach $\sim$3 $R_{J}$, this transient phase lasts less than $\sim$20 Myr.
The radii, when there are tides (e.g., for $e_{i}=0.40,0.30,0.20$), are close to one 
another simply because the final positions are close. 

However, we suggest that the rapid onset
of significant heating, followed by the muting of a significant effect on $R_p$
at later times, might be an artifact of our initial conditions $-$ such a combination 
of low $Q'_{p}$, high $e_i$, and small $a_i$ may not have been allowed to establish itself
without a more gradual feedback on the evolution that we, curiously, witness as an impulsive response.  Before 
$e_i$ would have been allowed to get that high or before $a_i$ would have been allowed to get that small,
tidal effects if $Q'_{p} = 10^5$ would no doubt have come into play.  This conclusion does, however, 
depend upon the processes, and their timescales, that result in high $e_i$ and in our initial $a_i$.
If early migration and eccentricity pumping can occur on very short timescales, an early,
almost impulsive, tidal response is possible.

\subsection{Coupling Enhanced Atmospheric Opacity with Tidal Heating}
\label{subsec:atm_opacity}

If one increases the atmospheric opacity, due either to an increased metallicity or 
to the possible effects of photolysis and/or non-equilibrium chemistry, the planet
retains heat better and can maintain a larger radius longer.  This effect was previously explored
by \citet{burrows_et_al2007}, who concluded that an enhanced opacity delays radius
shrinkage and can lead to larger EGP radii at a given age. They did their study
for fixed circular orbits and presented results for solar and 10$\times$solar 
atmospheric opacity. \citet{Liu_et_al_2008} also examined this effect, but 
at solar, 3$\times$solar, and 10$\times$solar opacities and explored the 
potential effects of tidal heating (at fixed, non-zero eccentricity).  
However, both studies kept the semi-major axis fixed. 
Here, we investigate the effect of enhanced opacity, but include orbit evolution as in 
\S\ref{subsec:typical_case} and \S\ref{subsec:strong_tides}.  For specificity, we focus 
on atmospheric opacities equivalent to those for chemical equilibrium abundances at 
solar and 3$\times$solar metallicity.  

Figure~\ref{fig:ms_evolution_fig3} depicts the simultaneous evolution of the planet's 
radius $R_{p}(t)$, eccentricity $e(t)$, semi-major axis $a(t)$, 
and power ratio $(\dot{E}_{\rm tide} /  \dot{E}_{\rm insolation})[t]$ for 
both solar (solid) and 3$\times$solar (dashed) atmospheric opacity.  We present our 
results for an illustrative subset of eccentricities ($e_{i}=0.60,0.55,0.40$),
ignore tides raised on the star, and use the same parameters, $Q'_{p}=10^{6.5}$ and $a_{i}=0.075$~AU,
employed in \S\ref{subsec:typical_case}. The black curves are reference models 
with no tides and $a=a_{i}=0.075$~AU.
The black, dashed curve for 3$\times$solar with no tidal effects 
recapitulates the results already obtained by \citet{burrows_et_al2007}. 
At the age of $\sim$1 Gyr, the radius increase effect is 5\%. 

The case for $e_{i}=0.60$ and with tidal effects is particularly illustrative
\footnote{As Fig. \ref{fig:ms_evolution_fig3} demonstrates, the 
qualitative behavior for different values of $e_{i}$ is the same.}.  With 
larger atmospheric opacity, the peaks in the radius and power ratio during 
the transient phase increase in magnitude, narrow, and shift 
towards younger ages.  Note that when tidal effects are included in the comparison
of the consequences of enhanced opacity, the results are not simply monotonic.
For $e_{i}=0.60$, an enhanced opacity leads first to a larger radius for ages up to $\sim$0.9 Gyr, 
then to a smaller radius for ages up to $\sim$2.8 Gyr, after which the planetary radius is again larger.
Importantly, the higher the atmospheric opacity, the faster the eccentricity evolves to zero
and the faster the semi-major axis tends to its final value.  This is a consequence 
of the larger values of $R_p$ possible with enhanced opacity and the stiff
dependence of $\dot{E}_{\rm tide}$ on $R_p$.  Interestingly, the 
final value of $a$ is the same as that reached for the lower atmospheric opacity. 
Why do we end up with the same final equilibrium circular orbit, as indicated
in Fig. \ref{fig:ms_evolution_fig3}, whereas in \S\ref{subsec:typical_case} different 
values of $e_{i}$ resulted in different final orbits? This behavior can be derived directly
from eqs. (\ref{eq:e}), (\ref{eq:a}), (\ref{eq:Etides}), and (\ref{eq:Etides_Einsol}).
For a given planet/star pair and the same values of $Q'_{p}$, $e_{i}$, and $a_{i}$,  
the amount of orbital energy transferred to the planet's interior is the same.
The increase in the atmospheric opacity decreases the rate of heat loss and the 
planet's  radius remains larger, longer. But this increase does not involve 
any transfer of energy inside the planet's interior. This is why for identical 
initial orbital energies, the final orbital energies and, therefore, the final 
semi-major axes are the same.  Angular momentum conservation can also be invoked 
to help explain this result.  Hence, when tidal effects subside, the final radius 
with enhanced atmospheric opacity is indeed larger than the one obtained with 
solar opacity, but the same orbital parameters, all else being equal. 
However, as Fig. \ref{fig:ms_evolution_fig3} indicates, the transient increase 
in $R_p$ when the atmospheric opacity is larger can be quite dramatic.

\subsection{The Particular Role of Tides Raised on the Star}
\label{subsec:tides_star}

Now, we examine what might happen when we include tidal dissipation in the star and set 
$Q'_{\ast}$ equal to various non-trivial values.  Figure~\ref{fig:ms_evolution_fig4} depicts the 
simultaneous evolution of $R_{p}(t)$, $e(t)$, $a(t)$, and $(\dot{E}_{\rm tide}/\dot{E}_{\rm insolation})[t]$
for $Q'_{p}=10^{6.5}$, $a_{i}=0.075$~AU, and solar-metallicty opacities, but this time for 
three different values of $Q'_{\ast}$: $10^{6}$ (dashed), $10^{5.5}$ (dotted), and $\infty$ (solid).
We replot from Fig. \ref{fig:ms_evolution_fig1} the three cases with no stellar tides, but 
with $e_{i}=0.60,0.55,0.40$, $Q'_{p}=10^{6.5}$, and $a=a_{i}=0.075$~AU.  
The solid black curve is for the case with no tides at all and $a=a_{i}=0.075$~AU.

Let us focus our discussion on the three cases with $e_{i}$~=~0.60, but 
for different values of $Q'_{\ast}$.  Including stellar tides, the lower 
the value of $Q'_{\ast}$, the higher, narrower, and earlier are the peaks 
of the radius and power ratio during the early transient phase. This behavior is similar to that encountered 
in \S\ref{subsec:atm_opacity} with enhanced opacity or in \S\ref{subsec:typical_case}
with higher initial eccentricity.  This time, however, it is caused by 
the additional terms involving $Q'_{\ast}$ in eqs. (\ref{eq:e}) and (\ref{eq:a}).
As seen in Fig.~\ref{fig:ms_evolution_fig4}, $e$ and $a$ decrease faster, but at 
a combined rate such that the ratios, $e^{2}/a^{11/2}$ in $\dot{E}_{\rm tide}/\dot{E}_{\rm insolation}$ (eq. \ref{eq:Etides_Einsol})
and $e^{2}/a^{15/2}$ in $\dot{E}_{\rm tide}$ (eq. \ref{eq:Etides}), are bigger 
than in the case without stellar tides. Therefore, more orbital energy is 
dissipated faster in the planet's interior.  As noted in \S\ref{subsec:typical_case} 
and \S\ref{subsec:atm_opacity}, the evolution is quite non-monotonic and nonlinear.  
Moreover, even when there is no pronounced peak, as for $e_{i}$~=~0.40, though the evolution is 
smoother, the general trends and behavior are similar. 

The case of $e_{i}=0.60$ and $Q'_{\ast}=10^{5.5}$ is particularly interesting. 
Figure \ref{fig:ms_evolution_fig4} shows that even after the orbit has already 
circularized ($e=0$), the semi-major axis, instead of stabilizing at a constant value, 
continues to decrease. In fact, the planet starts to spiral inward at an accelerating
rate, is eventually tidally disrupted, and collides with the star.  Equation 
(\ref{eq:a_when_e_equal_0}) in \S\ref{sec:model} gives the analytical evolution of $a$ 
after circularization, ignoring tidal disruption.   For $Q'_{\ast}=10^{5.5}$ 
and $e_{i}=0.60$, we intentionally stopped the evolution at $a=0.02$~AU, 
which is at roughly the Roche limit for this system: $\sim$$2.5 \left( \rho_{\ast}/\rho_{p} \right)^{1/3} R_{\ast}$, where $\rho_{\ast}$ 
and $\rho_{p}$ are the average densities of the star and the planet, respectively. 
While the planet is spiraling in, Fig.~\ref{fig:ms_evolution_fig4} shows that 
$R_{p}$ starts to shoot up.

This late-time behavior after the semi-axis has achieved small values and the orbital eccentricity is $\sim$zero 
has been seen before by \citet{rasio_1996}, \citet{Levrard_et_al_2009}, \citet{Jackson_et_al_2009}, and \citet{miller2009}.
However, whether such a dramatic effect would obtain depends sensitively on the value of $Q'_{\ast}$ 
(Fig. \ref{fig:ms_evolution_fig4}) and on whether tidal effects in the planet can synergistically force $a$ 
to achieve such low values that tidal dissipation in the star, which does not depend upon the eccentricity, 
can take over on stellar evolutionary timescales.  It would seem unlikely that the current sample of 
transiting and close-in EGPs are those for which we are just now catching a last glimpse 
before they are eaten by the star.  The implication for the parent population of exoplanets 
would appear extreme.  However, at this stage, we cannot eliminate the outside possibility
that, for a subset of very close-in EGPs, stellar tides might be important on timescales
short compared with their corresponding stellar ages.

\section{Application to HD~209458b}
\label{sec:HD209458b}

The previous sections were devoted to investigating the generic behavior 
of $R_p$ and orbital parameters when tidal dissipation in either the planet
or the star is at work.  Our main result is the emergence for a range of 
values of $e_i$ and $a_i$ of a transient phase of radius inflation 
which temporarily interrupts radius shrinkage. This phenomenon might explain
the larger-than-otherwise-expected planetary radii of an interesting
subset of the close-in transiting EGP population, either as a vestige of 
an epoch of earlier tidal heating, resetting the evolutionary clock \citep{Liu_et_al_2008}, 
or as a current episode of significant tidal heating.
An objective is to find a set of realistic values of $Q'_{p}$, $Q'_{\ast}$, $e_{i}$, and $a_{i}$
for which the currently measured values of $R_p$, $e$, and $a$ can be simultaneously explained.
Other factors, such as the atmospheric opacity, the possible presence of a dense core,
and the effect of a large heavy-element burden in the planetary envelope, also come into play,
and custom fits for each EGP are necessary. 

In this section, as an example we perform such an exercise 
for HD~209458b.  However, the general procedures and the generic 
features apply to all large-radius EGPs (e.g., TrES-4, WASP-12b, and WASP-6b) for which
transient tidal effects such as we have found in this paper are promising solutions. 
The observational data we use for HD 209458b are given in Table \ref{tab:HD209458_data}. 
Figure \ref{fig:ms_evolution_fig5} depicts solar-opacity model evolutions of 
HD 2094578b's radius $R_{p}(t)$, eccentricity $e(t)$, semi-major axis $a(t)$, 
and power ratio $(\dot{E}_{\rm tide} /  \dot{E}_{\rm insolation})[t]$.  We 
superpose on Fig. \ref{fig:ms_evolution_fig5} the measured values of $R_{p}$, $e$, 
and $a$, with their error boxes.  Also included on Fig. \ref{fig:ms_evolution_fig5},
but only in the top left panel, are four curves as reference cases with no tides. 
The dotted curve is for $a=a_{i}=0.085$~AU.  The other three non-solid 
curves are for solar, 3$\times$solar, and 10$\times$solar atmospheric opacities 
and for $a=0.047$~AU, the measured value of HD 209458b's semi-major axis. 

To fit the observations of HD 209458b, we have explored the effects of varying 
$Q'_{p}$ and $Q'_{\ast}$ from $10^{5}$ to $10^{8}$, $a_{i}$ from $0.10$~AU to $0.055$~AU, 
and $e_{i}$ from $0.80$ to $0.20$.  As a red line on Fig.~\ref{fig:ms_evolution_fig5},
we identify the model with the best-fitting parameter set, obtained by trial and error.
The associated parameters are: $Q'_{p}=10^{6.55}$, $Q'_{\ast}=10^{7.0}$, $a_{i}=0.085$~AU,
and $e_{i}=0.77$. To illustrate the extreme sensitivity to model parameters, 
we plot curves for $e_{i}$ ranging from 0.79 to 0.72, in steps of 0.01.  
Thus, by invoking this set of initial and planetary parameters and evolving 
the coupled suite of equations for $R_p$, $e$, and $a$, we can simultaneously 
explain the measured radius, eccentricity, and semi-major axis of HD 209458b.
In \S\ref{subsec:typical_case}, we described the sensitivity of 
the evolution to $e_{i}$. Indeed, when $e_{i}$ decreases, 
the transient peaks of the radius and power ratio widen, their height 
diminishes, and the epoch of transient heating shifts to older ages. 
The semi-major axis decreases more slowly and settles at higher values. 
With $Q'_{\ast} = 10^{7.0}$, the tides raised on the star are not strong enough 
to lead to inspiral of HD~209458b into its host star on a timescale less than 6 Gyr. 
However, for $Q'_{\ast} = 10^{7.0}$ there is still a slight decrease of the 
semi-major axis during the 14-Gyr duration of the simulations (not shown).  
This demonstrates that, for this choice of parameters, the orbit of 
HD 209458b is effectively stable.  For comparison, we have verified 
that values of $Q'_{\ast}$  of $10^{6.0}$ and $10^{5.5}$ lead to inspiral 
of HD~209458b. However,  the fit to the measurements is not as good.  This 
example demonstrates how sensitive the results are to the values of the chosen parameters.
Clearly, stronger independent constraints on these tidal $Q'$s are desirable.

Since the paper of \citet{burrows_et_al2007}, \citet{Torres_et_al_2008} have 
reappraised the age of HD 209458b, from $5.5^{+1.5}_{-1.5}$ to $3.1^{+0.8}_{-0.7}$ Gyr. 
As a consequence, as Fig.~\ref{fig:ms_evolution_fig5} shows, the curves for 3$\times$solar and 
10$\times$solar, but without tidal heating, now intercept the error box. 
We can in principle find a better fit if we choose an opacity between 
3 and 10$\times$solar, but still without invoking tidal effects.
Having said that, the introduction of tidal effects enables us
to fit HD 209458b's radius easily even with solar-opacity atmospheres.
In addition, we can now in principle provide a reasonable explanation for its 
observed eccentricity (and/or upper limit) and semi-major axis.  Moreover, if one  
needed to invoke high-metallicity (not just high-opacity) atmospheres 
to explain HD 209458b's radius and, thereby high-metallicity 
envelopes, a transient tidal phase with significant heating now 
allows even the shrinking effect of heavy elements in the envelope 
to be compensated for.  This system is a good example of a planet whose 
orbit has almost circularized, but that still has an inflated radius.  
Hence, as we have suggested in this subsection and in \S\ref{subsec:typical_case}, 
the radius of HD 209458b may be due to the former action of 
tides during an earlier epoch of non-circular evolution.  
This explanation might also be germane to the TrES-4, WASP-12b, 
and WASP-6b planets, as well as to other large-radius EGPs.

There exists an interesting, if counterintuitive, systematic behavior 
in models for the simultaneous evolution of $a$, $e$, and $R_{P}$.
For a given final position ($a,e$), an initially
higher $e_i$ does not necessarily result in a larger radius.
The radius evolution is a strongly non-linear function of the initial
conditions and as Fig. \ref{fig:ms_evolution_fig1} indicates, different $e_i$s 
result in different radius evolutionary curves that intersect. Comparing radii for models 
with different initial values of $a_i$ and $e_i$ requires one to specify the age of the planet. 
Among the 7 curves on Fig. \ref{fig:ms_evolution_fig5} (for $e_i$=0.72 to 0.79), only
those for $e_i = 0.77$ and $= 0.76$ lie within the $a$, $e$, age, and $R_{p}$ error boxes.
Moreover, and perhaps counterintuitively, at the age of 
HD 209458b the radius for $e_i$=0.77 is smaller than the radius for
$e_i$=0.76.

Hence, a model with a higher initial eccentricity peaks at a higher tidal
heating rate, but at earlier ages. At the later times at which models
with lower initial eccentricity peak, they have higher heating rates
than the former at that same time. In the same vein, for the same
$e_i$, starting with lower $a_i$ results in a higher tidal heating rate
at earlier ages than the tidal heating rate that eventually obtains
when starting with a larger $a_i$.

\section{Conclusions and Discussion}
\label{sec:conclusion}

In this paper, we have found that if an EGP after early migration and dynamical 
evolution is left in a tight orbit ($a_i$ $\sles$ $0.2$~AU) with 
a modest to high eccentricity ($\gtrsim$${0.2}$), its subsequent evolution
due to tidal dissipation can qualitatively alter our interpretation of 
its measured radius.  Using a formalism in which the planet's radius and
orbit are consistently and simultaneously evolved, we have found that a transient 
phase of rapid tidal heating and radius expansion can help explain the large
radii measured for some transiting EGPs even after this phase has subsided.
This explanation is straighforward for HD 209458b, but it might also be a 
factor in the large radii observed for, for example, TrES-4 \citep{Mandushev_et_al_2007,Torres_et_al_2008,Sozzetti_et_al_2008}, 
WASP-12b \citep{Hebb_et_al_2009}, WASP-6b \citep{Gillon_et_al_2009_2}, and WASP-4b 
\citep{Wilson_et_al_2008,Gillon_et_al_2009_1,Winn_et_al_2009_1}.

We parameterized our models using a range of planet and star tidal dissipation factors, 
a range of initial values of the eccentricity and semi-major axis, and three realizations 
of the planet's atmospheric opacity. Our main conclusions are:
\begin{itemize}\itemsep0cm
\item A giant planet's radius can undergo a transient phase of inflation due to tides that temporarily interrupts 
      its shrinkage and resets its evolutionary clock. The upshot is that, for suitable parameters 
      (e.g., $Q'_{p}$ and $Q'_{\ast}$ of $10^{5.0}$--$10^{7.0}$, $a_i$ less than $\sim$0.2 AU, and $e_i \gtrsim 0.2$), $R_p$
      is measurably larger than it otherwise would have been, even after tidal heating has subsided.
      We have demonstrated that a planet whose orbit has circularized can still have an inflated radius 
      due to the former action of tides.
\item Extremely strong tidal heating in a planet will fade early in its life. Under such circumstances, the 
      effect on $R_p$ will be negligible at the typical ages of the observed transiting planets.
\item Higher atmospheric opacities can enhance and accelerate the transient phase of radius inflation and accelerate orbital evolution.  
\item The tides raised on the star also enhance and accelerate the transient phase of radius inflation. However, even after
      the orbit circularizes due to tidal dissipation in the planet, the orbit does not stabilize $-$ for small enough values of
      $Q'_{\ast}$, the planet can plunge inward, and is then tidally disrupted and consumed by the star.
\item Radius and orbit evolution are strongly non-linear and stiff functions of the 
      parameters $Q'_{p}$, $Q'_{\ast}$, $e_{i}$, $a_{i}$, and atmospheric opacity.
      Custom fits to each planet/star system, rather than pre-calculated look-up tables, are to be preferred.   
\item The parameters ($Q'_{p}$, $Q'_{\ast}$,$e_{i}$, $a_{i}$)$=$($10^{6.55},10^{7.0},0.77,0.085~\rm AU$), with 
      a solar atmospheric opacity, provide a satisfactory fit to the measured radius of HD 209458b, while also
      being consistent with its current orbit. In general, the higher the atmospheric opacity
      the less tidal heating needs to be invoked to fit its observed $R_{P}$.    
\item Given its new, younger age, the measured radius of HD 209458b can now also be fit without tides, but only if 
      the atmospheric opacity is similar to those with equilibrium chemical 
      abundances for metallicities greater than $\sim$3$\times$solar. However, 
      unless the atmospheric abundances can be measured, it will be
      difficult to distinguish the different predictions, particularly given the
      current ambiguity in $Q'$.  Nevertheless, if there are independent (theoretical?)
      constraints on $Q'$, and/or precision measurements of the atmospheric spectra,
      one might be able to discriminate between the two different explanations (vestigal 
      tidal effects and higher atmospheric opacity).
\end{itemize}

There are numerous caveats that should be borne in mind when considering our results:  

\begin{enumerate}
\item Our atmospheric boundary conditions assume that day side and night side cooling are the same.  
Credible general circulation models (GCMs) \citep{Goodman_2008,showman_et_al2008b,showman_et_al2008,
showman_et_al_2007,dobbs-dixon+lin2008,langton+laughlin2008,cho_et_al2008,menou+rauscher2008} might address this issue most usefully.
\item We have fixed the properties of the host star during our integrations.  
On timescales of a few Gyr, the evolution of the star might be relevant.  
\item The tidal dissipation factors $Q'$ are very poorly constrained and are free parameters of 
our models.  A firmer theoretical handle on their values and dependences would be very useful
\citep{Ogilvie_and_Lin_2004, Wu_2005_1,Wu_2005_2,Goodman+Lackner_2008}.
\item We have assumed that the planet's spin is synchronized (tidally-locked) and 
have ignored the spin angular momentum of both the planet and the star.
This is generally a good approximation, but there may be exceptional cases.
\item We have assumed that the tidal heating power is deposited solely in the convective zone, in which it is rapidly mixed.  
If much of the heat is instead deposited in the radiative atmosphere, as might be expected if circularization
is dominated by Hough modes \citep{Ogilvie_and_Lin_2007}, our results would need alteration.  
\item We have neglected the obliquities of both the star and the planet. 
\item We start our calculations after early disk evolution and dispersal, and assume that the anomalously large
planets, for which we might invoke tidal effects, are left after these early phases with high eccentricities and 
small semi-major axes. These assumptions are at present almost wholly unconstrained.  We may, however, be
able to turn the question around and someday use the current observations to constrain an earlier episode of vigorous 
tidal heating.
\end{enumerate}

We have demonstrated that, given the right parameters and initial conditions, early tidal heating can result in large EGP radii, even
while leaving the planet's current orbital parameters at modest values for which tidal heating 
is not expected to be important. These vestigal effects of an earlier transient phase might echo 
into the present to help explain the anomalously large radii of a small, but interesting, subset
of transiting extrasolar giant planets.  Our formalism can easily be applied to other 
inflated transiting EGPs, such as TrES-4 \citep{Mandushev_et_al_2007,Torres_et_al_2008,Sozzetti_et_al_2008}, 
WASP-12b \citep{Hebb_et_al_2009}, WASP-6b \citep{Gillon_et_al_2009_2}, and WASP-4b \citep{Wilson_et_al_2008,Gillon_et_al_2009_1,Winn_et_al_2009_1}, 
and we plan to do so in the near future.


\acknowledgements
We thank Ivan Hubeny for help with boundary condition issues, and Brian Jackson for instructive insights into  
the tidal evolution equations. We also thank Dave Spiegel, Jeremy Goodman, Rosemary Mardling, 
and Roman Rafikov for helpful discussions and an anonymous referee for suggesting changes that
materially improved the manuscript. This study was supported by NASA grant 
NNX07AG80G and under JPL/Spitzer Agreements 1328092, 1348668, and 1312647.

\bibliography{biblio}

\begin{thebibliography}{97}
\expandafter\ifx\csname natexlab\endcsname\relax\def\natexlab#1{#1}\fi

\bibitem[{{Bakos} {et~al.}(2007){Bakos}, {Noyes}, {Kov{\'a}cs}, {Latham},
  {Sasselov}, {Torres}, {Fischer}, {Stefanik}, {Sato}, {Johnson}, {P{\'a}l},
  {Marcy}, {Butler}, {Esquerdo}, {Stanek}, {L{\'a}z{\'a}r}, {Papp}, {S{\'a}ri},
  \& {Sip{\H o}cz}}]{bakos_et_al2007a}
{Bakos}, G.~{\'A}., {Noyes}, R.~W., {Kov{\'a}cs}, G., {Latham}, D.~W.,
  {Sasselov}, D.~D., {Torres}, G., {Fischer}, D.~A., {Stefanik}, R.~P., {Sato},
  B., {Johnson}, J.~A., {P{\'a}l}, A., {Marcy}, G.~W., {Butler}, R.~P.,
  {Esquerdo}, G.~A., {Stanek}, K.~Z., {L{\'a}z{\'a}r}, J., {Papp}, I.,
  {S{\'a}ri}, P., \& {Sip{\H o}cz}, B. 2007, \apj, 656, 552

\bibitem[{{Baraffe} {et~al.}(2006){Baraffe}, {Alibert}, {Chabrier}, \&
  {Benz}}]{Baraffe_et_al_2006}
{Baraffe}, I., {Alibert}, Y., {Chabrier}, G., \& {Benz}, W. 2006, \aap, 450,
  1221

\bibitem[{{Baraffe} {et~al.}(2008){Baraffe}, {Chabrier}, \&
  {Barman}}]{Baraffe_et_al_2008}
{Baraffe}, I., {Chabrier}, G., \& {Barman}, T. 2008, \aap, 482, 315

\bibitem[{{Baraffe} {et~al.}(2003){Baraffe}, {Chabrier}, {Barman}, {Allard}, \&
  {Hauschildt}}]{Baraffe_et_al_2003}
{Baraffe}, I., {Chabrier}, G., {Barman}, T.~S., {Allard}, F., \& {Hauschildt},
  P.~H. 2003, \aap, 402, 701

\bibitem[{{Baraffe} {et~al.}(2005){Baraffe}, {Chabrier}, {Barman}, {Selsis},
  {Allard}, \& {Hauschildt}}]{Baraffe_et_al_2005}
{Baraffe}, I., {Chabrier}, G., {Barman}, T.~S., {Selsis}, F., {Allard}, F., \&
  {Hauschildt}, P.~H. 2005, \aap, 436, L47

\bibitem[{{Baraffe} {et~al.}(2004){Baraffe}, {Selsis}, {Chabrier}, {Barman},
  {Allard}, {Hauschildt}, \& {Lammer}}]{Baraffe_et_al_2004}
{Baraffe}, I., {Selsis}, F., {Chabrier}, G., {Barman}, T.~S., {Allard}, F.,
  {Hauschildt}, P.~H., \& {Lammer}, H. 2004, \aap, 419, L13

\bibitem[{{Barnes} {et~al.}(2009){Barnes}, {Jackson}, {Raymond}, {West}, \&
  {Greenberg}}]{Barnes_et_al_2009}
{Barnes}, R., {Jackson}, B., {Raymond}, S.~N., {West}, A.~A., \& {Greenberg},
  R. 2009, \apj, 695, 1006

\bibitem[{{Bodenheimer} {et~al.}(2003){Bodenheimer}, {Laughlin}, \&
  {Lin}}]{Bodenheimer_et_al_2003}
{Bodenheimer}, P., {Laughlin}, G., \& {Lin}, D.~N.~C. 2003, \apj, 592, 555

\bibitem[{{Bodenheimer} {et~al.}(2001){Bodenheimer}, {Lin}, \&
  {Mardling}}]{Bodenheimer_et_al_2001}
{Bodenheimer}, P., {Lin}, D.~N.~C., \& {Mardling}, R.~A. 2001, \apj, 548, 466

\bibitem[{{Burrows} {et~al.}(2000){Burrows}, {Guillot}, {Hubbard}, {Marley},
  {Saumon}, {Lunine}, \& {Sudarsky}}]{burrows_et_al2000}
{Burrows}, A., {Guillot}, T., {Hubbard}, W.~B., {Marley}, M.~S., {Saumon}, D.,
  {Lunine}, J.~I., \& {Sudarsky}, D. 2000, \apjl, 534, L97

\bibitem[{{Burrows} {et~al.}(1993){Burrows}, {Hubbard}, {Saumon}, \&
  {Lunine}}]{Burrows_et_al_1993}
{Burrows}, A., {Hubbard}, W.~B., {Saumon}, D., \& {Lunine}, J.~I. 1993, \apj,
  406, 158

\bibitem[{{Burrows} {et~al.}(2007){Burrows}, {Hubeny}, {Budaj}, \&
  {Hubbard}}]{burrows_et_al2007}
{Burrows}, A., {Hubeny}, I., {Budaj}, J., \& {Hubbard}, W.~B. 2007, \apj, 661,
  502

\bibitem[{{Burrows} {et~al.}(2004){Burrows}, {Hubeny}, {Hubbard}, {Sudarsky},
  \& {Fortney}}]{burrows_et_al2004}
{Burrows}, A., {Hubeny}, I., {Hubbard}, W.~B., {Sudarsky}, D., \& {Fortney},
  J.~J. 2004, \apjl, 610, L53

\bibitem[{{Burrows} {et~al.}(1997){Burrows}, {Marley}, {Hubbard}, {Lunine},
  {Guillot}, {Saumon}, {Freedman}, {Sudarsky}, \& {Sharp}}]{Burrows_et_al_1997}
{Burrows}, A., {Marley}, M., {Hubbard}, W.~B., {Lunine}, J.~I., {Guillot}, T.,
  {Saumon}, D., {Freedman}, R., {Sudarsky}, D., \& {Sharp}, C. 1997, \apj, 491,
  856

\bibitem[{{Burrows} {et~al.}(2003){Burrows}, {Sudarsky}, \&
  {Hubbard}}]{burrows_et_al2003}
{Burrows}, A., {Sudarsky}, D., \& {Hubbard}, W.~B. 2003, \apj, 594, 545

\bibitem[{{Chabrier} \& {Baraffe}(2007)}]{chabrier+baraffe2007}
{Chabrier}, G. \& {Baraffe}, I. 2007, \apjl, 661, L81

\bibitem[{{Chabrier} {et~al.}(2004){Chabrier}, {Barman}, {Baraffe}, {Allard},
  \& {Hauschildt}}]{Chabrier_et_al_2004}
{Chabrier}, G., {Barman}, T., {Baraffe}, I., {Allard}, F., \& {Hauschildt},
  P.~H. 2004, \apjl, 603, L53

\bibitem[{{Charbonneau} {et~al.}(2000){Charbonneau}, {Brown}, {Latham}, \&
  {Mayor}}]{charbonneau_et_al2000}
{Charbonneau}, D., {Brown}, T.~M., {Latham}, D.~W., \& {Mayor}, M. 2000, \apjl,
  529, L45

\bibitem[{{Chatterjee} {et~al.}(2008){Chatterjee}, {Ford}, {Matsumura}, \&
  {Rasio}}]{Chatterjee_et_al_2008}
{Chatterjee}, S., {Ford}, E.~B., {Matsumura}, S., \& {Rasio}, F.~A. 2008, \apj,
  686, 580

\bibitem[{{Cho} {et~al.}(2008){Cho}, {Menou}, {Hansen}, \&
  {Seager}}]{cho_et_al2008}
{Cho}, J.~Y.-K., {Menou}, K., {Hansen}, B.~M.~S., \& {Seager}, S. 2008, \apj,
  675, 817

\bibitem[{{Darwin}(1880)}]{Darwin_1880}
{Darwin}, G.~H. 1880, \nat, 21, 235

\bibitem[{{Dickey} {et~al.}(1994){Dickey}, {Bender}, {Faller}, {Newhall},
  {Ricklefs}, {Ries}, {Shelus}, {Veillet}, {Whipple}, {Wiant}, {Williams}, \&
  {Yoder}}]{Dickey_et_al_1994}
{Dickey}, J.~O., {Bender}, P.~L., {Faller}, J.~E., {Newhall}, X.~X.,
  {Ricklefs}, R.~L., {Ries}, J.~G., {Shelus}, P.~J., {Veillet}, C., {Whipple},
  A.~L., {Wiant}, J.~R., {Williams}, J.~G., \& {Yoder}, C.~F. 1994, Science,
  265, 482

\bibitem[{{Dobbs-Dixon} \& {Lin}(2008)}]{dobbs-dixon+lin2008}
{Dobbs-Dixon}, I. \& {Lin}, D.~N.~C. 2008, \apj, 673, 513

\bibitem[{{Dobbs-Dixon} {et~al.}(2004){Dobbs-Dixon}, {Lin}, \&
  {Mardling}}]{dobbs-dixon+lin2004}
{Dobbs-Dixon}, I., {Lin}, D.~N.~C., \& {Mardling}, R.~A. 2004, \apj, 610, 464

\bibitem[{{Eggleton} {et~al.}(1998){Eggleton}, {Kiseleva}, \&
  {Hut}}]{Eggleton_et_al_1998}
{Eggleton}, P.~P., {Kiseleva}, L.~G., \& {Hut}, P. 1998, \apj, 499, 853

\bibitem[{{Fabrycky} {et~al.}(2007){Fabrycky}, {Johnson}, \&
  {Goodman}}]{fabrycky_et_al2007}
{Fabrycky}, D.~C., {Johnson}, E.~T., \& {Goodman}, J. 2007, \apj, 665, 754

\bibitem[{{Ferraz-Mello} {et~al.}(2008){Ferraz-Mello}, {Rodr{\'{\i}}guez}, \&
  {Hussmann}}]{Ferraz-Mello_et_al_2008}
{Ferraz-Mello}, S., {Rodr{\'{\i}}guez}, A., \& {Hussmann}, H. 2008, Celestial
  Mechanics and Dynamical Astronomy, 101, 171

\bibitem[{{Ford} \& {Rasio}(2008)}]{Ford_and_Rasio_2008}
{Ford}, E.~B. \& {Rasio}, F.~A. 2008, \apj, 686, 621

\bibitem[{{Ford} {et~al.}(2003){Ford}, {Rasio}, \& {Yu}}]{Ford_et_al_2003}
{Ford}, E.~B., {Rasio}, F.~A., \& {Yu}, K. 2003, in Astronomical Society of the
  Pacific Conference Series, Vol. 294, Scientific Frontiers in Research on
  Extrasolar Planets, ed. D.~{Deming} \& S.~{Seager}, 181--188

\bibitem[{{Fortney} \& {Hubbard}(2004)}]{Fortney_and_Hubbard_2004}
{Fortney}, J.~J. \& {Hubbard}, W.~B. 2004, \apj, 608, 1039

\bibitem[{{Fortney} {et~al.}(2007){Fortney}, {Marley}, \&
  {Barnes}}]{fortney_et_al2007}
{Fortney}, J.~J., {Marley}, M.~S., \& {Barnes}, J.~W. 2007, \apj, 659, 1661

\bibitem[{{Fortney} {et~al.}(2008){Fortney}, {Marley}, {Saumon}, \&
  {Lodders}}]{Fortney_et_al_2008}
{Fortney}, J.~J., {Marley}, M.~S., {Saumon}, D., \& {Lodders}, K. 2008, \apj,
  683, 1104

\bibitem[{{Gavrilov} \& {Zharkov}(1977)}]{Gavrilov_and_Zharkov_1977}
{Gavrilov}, S.~V. \& {Zharkov}, V.~N. 1977, Icarus, 32, 443

\bibitem[{{Gillon} {et~al.}(2009{\natexlab{a}}){Gillon}, {Anderson}, {Triaud},
  {Hellier}, {Maxted}, {Pollaco}, {Queloz}, {Smalley}, {West}, {Wilson},
  {Bentley}, {Collier Cameron}, {Enoch}, {Hebb}, {Horne}, {Irwin}, {Joshi},
  {Lister}, {Mayor}, {Pepe}, {Parley}, {Segransan}, {Udry}, \&
  {Wheatley}}]{Gillon_et_al_2009_2}
{Gillon}, M., {Anderson}, D.~R., {Triaud}, A.~H.~M.~J., {Hellier}, C.,
  {Maxted}, P.~F.~L., {Pollaco}, D., {Queloz}, D., {Smalley}, B., {West},
  R.~G., {Wilson}, D.~M., {Bentley}, S.~J., {Collier Cameron}, A., {Enoch}, B.,
  {Hebb}, L., {Horne}, K., {Irwin}, J., {Joshi}, Y.~C., {Lister}, T.~A.,
  {Mayor}, M., {Pepe}, F., {Parley}, N., {Segransan}, D., {Udry}, S., \&
  {Wheatley}, P.~J. 2009{\natexlab{a}}, submitted to \aap~(arXiv:0901.4705)

\bibitem[{{Gillon} {et~al.}(2009{\natexlab{b}}){Gillon}, {Smalley}, {Hebb},
  {Anderson}, {Triaud}, {Hellier}, {Maxted}, {Queloz}, \&
  {Wilson}}]{Gillon_et_al_2009_1}
{Gillon}, M., {Smalley}, B., {Hebb}, L., {Anderson}, D.~R., {Triaud},
  A.~H.~M.~J., {Hellier}, C., {Maxted}, P.~F.~L., {Queloz}, D., \& {Wilson},
  D.~M. 2009{\natexlab{b}}, \aap, 496, 259

\bibitem[{{Goldreich} \& {Sari}(2003)}]{Goldreich_and_Sari_2003}
{Goldreich}, P. \& {Sari}, R. 2003, \apj, 585, 1024

\bibitem[{{Goldreich} \& {Soter}(1966)}]{Goldreich+Soter_1966}
{Goldreich}, P. \& {Soter}, S. 1966, Icarus, 5, 375

\bibitem[{{Goldreich}(1963)}]{Goldreich_1963}
{Goldreich}, R. 1963, \mnras, 126, 257

\bibitem[{{Goodman}(2008)}]{Goodman_2008}
{Goodman}, J. 2008, (arXiv:0810.1282)

\bibitem[{{Goodman} \& {Lackner}(2008)}]{Goodman+Lackner_2008}
{Goodman}, J. \& {Lackner}, C. 2008, submitted to \apj~(arXiv:0812.1028)

\bibitem[{{Gu} {et~al.}(2004){Gu}, {Bodenheimer}, \& {Lin}}]{Gu_et_al_2004}
{Gu}, P.-G., {Bodenheimer}, P.~H., \& {Lin}, D.~N.~C. 2004, \apj, 608, 1076

\bibitem[{{Gu} {et~al.}(2003){Gu}, {Lin}, \& {Bodenheimer}}]{Gu_et_al_2003}
{Gu}, P.-G., {Lin}, D.~N.~C., \& {Bodenheimer}, P.~H. 2003, \apj, 588, 509

\bibitem[{{Guillot} {et~al.}(1996){Guillot}, {Burrows}, {Hubbard}, {Lunine}, \&
  {Saumon}}]{guillot_et_al1996}
{Guillot}, T., {Burrows}, A., {Hubbard}, W.~B., {Lunine}, J.~I., \& {Saumon},
  D. 1996, \apjl, 459, L35+

\bibitem[{{Guillot} {et~al.}(2006){Guillot}, {Santos}, {Pont}, {Iro}, {Melo},
  \& {Ribas}}]{guillot06}
{Guillot}, T., {Santos}, N.~C., {Pont}, F., {Iro}, N., {Melo}, C., \& {Ribas},
  I. 2006, \aap, 453, L21

\bibitem[{{Guillot} \& {Showman}(2002)}]{guillot+showman2002}
{Guillot}, T. \& {Showman}, A.~P. 2002, \aap, 385, 156

\bibitem[{{Hebb} {et~al.}(2009){Hebb}, {Collier-Cameron}, {Loeillet},
  {Pollacco}, {H{\'e}brard}, {Street}, {Bouchy}, {Stempels}, {Moutou},
  {Simpson}, {Udry}, {Joshi}, {West}, {Skillen}, {Wilson}, {McDonald},
  {Gibson}, {Aigrain}, {Anderson}, {Benn}, {Christian}, {Enoch}, {Haswell},
  {Hellier}, {Horne}, {Irwin}, {Lister}, {Maxted}, {Mayor}, {Norton}, {Parley},
  {Pont}, {Queloz}, {Smalley}, \& {Wheatley}}]{Hebb_et_al_2009}
{Hebb}, L., {Collier-Cameron}, A., {Loeillet}, B., {Pollacco}, D.,
  {H{\'e}brard}, G., {Street}, R.~A., {Bouchy}, F., {Stempels}, H.~C.,
  {Moutou}, C., {Simpson}, E., {Udry}, S., {Joshi}, Y.~C., {West}, R.~G.,
  {Skillen}, I., {Wilson}, D.~M., {McDonald}, I., {Gibson}, N.~P., {Aigrain},
  S., {Anderson}, D.~R., {Benn}, C.~R., {Christian}, D.~J., {Enoch}, B.,
  {Haswell}, C.~A., {Hellier}, C., {Horne}, K., {Irwin}, J., {Lister}, T.~A.,
  {Maxted}, P., {Mayor}, M., {Norton}, A.~J., {Parley}, N., {Pont}, F.,
  {Queloz}, D., {Smalley}, B., \& {Wheatley}, P.~J. 2009, \apj, 693, 1920

\bibitem[{{Hubeny} \& {Lanz}(1995)}]{Hubeny_and_Lanz_1995}
{Hubeny}, I. \& {Lanz}, T. 1995, \apj, 439, 875

\bibitem[{{Hut}(1981)}]{Hut_1981}
{Hut}, P. 1981, \aap, 99, 126

\bibitem[{{Jackson} {et~al.}(2008{\natexlab{a}}){Jackson}, {Barnes}, \&
  {Greenberg}}]{Jackson_et_al_2008_4}
{Jackson}, B., {Barnes}, R., \& {Greenberg}, R. 2008{\natexlab{a}}, \mnras,
  391, 237

\bibitem[{{Jackson} {et~al.}(2008{\natexlab{b}}){Jackson}, {Greenberg}, \&
  {Barnes}}]{Jackson_et_al_2008_1}
{Jackson}, B., {Greenberg}, R., \& {Barnes}, R. 2008{\natexlab{b}}, in IAU
  Symposium, Vol. 249, IAU Symposium, 187--196

\bibitem[{{Jackson} {et~al.}(2008{\natexlab{c}}){Jackson}, {Greenberg}, \&
  {Barnes}}]{Jackson_et_al_2008_2}
{Jackson}, B., {Greenberg}, R., \& {Barnes}, R. 2008{\natexlab{c}}, \apj, 678,
  1396

\bibitem[{{Jackson} {et~al.}(2008{\natexlab{d}}){Jackson}, {Greenberg}, \&
  {Barnes}}]{Jackson_et_al_2008_3}
---. 2008{\natexlab{d}}, \apj, 681, 1631

\bibitem[{{Jackson} {et~al.}(2009){Jackson}, {Greenberg}, \&
  {Barnes}}]{Jackson_et_al_2009}
{Jackson}, B., {Greenberg}, R., \& {Barnes}, R. 2009, in American Astronomical
  Society Meeting Abstracts, Vol. 213, American Astronomical Society Meeting
  Abstracts, 351.01

\bibitem[{{Johns-Krull} {et~al.}(2008){Johns-Krull}, {McCullough}, {Burke},
  {Valenti}, {Janes}, {Heasley}, {Prato}, {Bissinger}, {Fleenor}, {Foote},
  {Garcia-Melendo}, {Gary}, {Howell}, {Mallia}, {Masi}, \&
  {Vanmunster}}]{Johns-Krull_et_al_2008}
{Johns-Krull}, C.~M., {McCullough}, P.~R., {Burke}, C.~J., {Valenti}, J.~A.,
  {Janes}, K.~A., {Heasley}, J.~N., {Prato}, L., {Bissinger}, R., {Fleenor},
  M., {Foote}, C.~N., {Garcia-Melendo}, E., {Gary}, B.~L., {Howell}, P.~J.,
  {Mallia}, F., {Masi}, G., \& {Vanmunster}, T. 2008, \apj, 677, 657

\bibitem[{{Johnson} {et~al.}(2008){Johnson}, {Winn}, {Narita}, {Enya},
  {Williams}, {Marcy}, {Sato}, {Ohta}, {Taruya}, {Suto}, {Turner}, {Bakos},
  {Butler}, {Vogt}, {Aoki}, {Tamura}, {Yamada}, {Yoshii}, \&
  {Hidas}}]{Johnson_et_al_2008}
{Johnson}, J.~A., {Winn}, J.~N., {Narita}, N., {Enya}, K., {Williams},
  P.~K.~G., {Marcy}, G.~W., {Sato}, B., {Ohta}, Y., {Taruya}, A., {Suto}, Y.,
  {Turner}, E.~L., {Bakos}, G., {Butler}, R.~P., {Vogt}, S.~S., {Aoki}, W.,
  {Tamura}, M., {Yamada}, T., {Yoshii}, Y., \& {Hidas}, M. 2008, \apj, 686, 649

\bibitem[{{Juri{\'c}} \& {Tremaine}(2008)}]{Juric_and_Tremaine_2008}
{Juri{\'c}}, M. \& {Tremaine}, S. 2008, \apj, 686, 603

\bibitem[{{Kaula}(1968)}]{Kaula_1968}
{Kaula}, W.~M. 1968, {An introduction to planetary physics - The terrestrial
  planets} (Space Science Text Series, New York: Wiley, 1968)

\bibitem[{{Knutson} {et~al.}(2007){Knutson}, {Charbonneau}, {Noyes}, {Brown},
  \& {Gilliland}}]{knutson_et_al2007a}
{Knutson}, H.~A., {Charbonneau}, D., {Noyes}, R.~W., {Brown}, T.~M., \&
  {Gilliland}, R.~L. 2007, \apj, 655, 564

\bibitem[{{Kurucz}(1994)}]{Kurucz_1994}
{Kurucz}, R. 1994, Solar abundance model atmospheres for 0,1,2,4,8 km/s.~Kurucz
  CD-ROM No.~19.~ Cambridge, Mass.: Smithsonian Astrophysical Observatory,
  1994., 19

\bibitem[{{Langton} \& {Laughlin}(2008)}]{langton+laughlin2008}
{Langton}, J. \& {Laughlin}, G. 2008, \apj, 674, 1106

\bibitem[{{Laughlin} {et~al.}(2005){Laughlin}, {Wolf}, {Vanmunster},
  {Bodenheimer}, {Fischer}, {Marcy}, {Butler}, \& {Vogt}}]{laughlin_et_al2005}
{Laughlin}, G., {Wolf}, A., {Vanmunster}, T., {Bodenheimer}, P., {Fischer}, D.,
  {Marcy}, G., {Butler}, P., \& {Vogt}, S. 2005, \apj, 621, 1072

\bibitem[{{Levrard} {et~al.}(2007){Levrard}, {Correia}, {Chabrier}, {Baraffe},
  {Selsis}, \& {Laskar}}]{levrard_et_al2007}
{Levrard}, B., {Correia}, A.~C.~M., {Chabrier}, G., {Baraffe}, I., {Selsis},
  F., \& {Laskar}, J. 2007, \aap, 462, L5

\bibitem[{{Levrard} {et~al.}(2009){Levrard}, {Winisdoerffer}, \&
  {Chabrier}}]{Levrard_et_al_2009}
{Levrard}, B., {Winisdoerffer}, C., \& {Chabrier}, G. 2009, \apjl, 692, L9

\bibitem[{{Liu} {et~al.}(2008){Liu}, {Burrows}, \& {Ibgui}}]{Liu_et_al_2008}
{Liu}, X., {Burrows}, A., \& {Ibgui}, L. 2008, \apj, 687, 1191

\bibitem[{{Love}(1927)}]{Love_1927}
{Love}, A.~E.~H. 1927, {A Treatise on the Mathematical Theory of Elasticity}
  (Dover, New York)

\bibitem[{{Madhusudhan} \& {Winn}(2009)}]{Madhusudhan_and_Winn_2008}
{Madhusudhan}, N. \& {Winn}, J.~N. 2009, \apj, 693, 784

\bibitem[{{Mandushev} {et~al.}(2007){Mandushev}, {O'Donovan}, {Charbonneau},
  {Torres}, {Latham}, {Bakos}, {Dunham}, {Sozzetti}, {Fern{\'a}ndez},
  {Esquerdo}, {Everett}, {Brown}, {Rabus}, {Belmonte}, \&
  {Hillenbrand}}]{Mandushev_et_al_2007}
{Mandushev}, G., {O'Donovan}, F.~T., {Charbonneau}, D., {Torres}, G., {Latham},
  D.~W., {Bakos}, G.~{\'A}., {Dunham}, E.~W., {Sozzetti}, A., {Fern{\'a}ndez},
  J.~M., {Esquerdo}, G.~A., {Everett}, M.~E., {Brown}, T.~M., {Rabus}, M.,
  {Belmonte}, J.~A., \& {Hillenbrand}, L.~A. 2007, \apjl, 667, L195

\bibitem[{{Mardling}(2007)}]{Mardling_2007}
{Mardling}, R.~A. 2007, \mnras, 382, 1768

\bibitem[{{Mardling} \& {Lin}(2002)}]{Mardling_and_Lin_2002}
{Mardling}, R.~A. \& {Lin}, D.~N.~C. 2002, \apj, 573, 829

\bibitem[{{Mardling} \& {Lin}(2004)}]{Mardling_and_Lin_2004}
---. 2004, \apj, 614, 955

\bibitem[{{Marley} {et~al.}(2007){Marley}, {Fortney}, {Hubickyj},
  {Bodenheimer}, \& {Lissauer}}]{Marley_et_al_2007}
{Marley}, M.~S., {Fortney}, J.~J., {Hubickyj}, O., {Bodenheimer}, P., \&
  {Lissauer}, J.~J. 2007, \apj, 655, 541

\bibitem[{{Menou} \& {Rauscher}(2008)}]{menou+rauscher2008}
{Menou}, K. \& {Rauscher}, E. 2008, submitted to \apj~(arXiv:0809.1671)

\bibitem[{{Miller} {et~al.}(2009){Miller}, {Fortney}, \&
  {Jackson}}]{miller2009}
{Miller}, N., {Fortney}, J., \& {Jackson}, B. 2009, B.A.A.S. 402.07

\bibitem[{{Murray} \& {Dermott}(1999)}]{Murray_et_Dermott_1999}
{Murray}, C.~D. \& {Dermott}, S.~F. 1999, {Solar system dynamics} (Solar system
  dynamics by Murray, C.~D., 1999)

\bibitem[{{Nagasawa} {et~al.}(2008){Nagasawa}, {Ida}, \&
  {Bessho}}]{Nagasawa_et_al_2008}
{Nagasawa}, M., {Ida}, S., \& {Bessho}, T. 2008, \apj, 678, 498

\bibitem[{{Ogilvie} \& {Lin}(2004)}]{Ogilvie_and_Lin_2004}
{Ogilvie}, G.~I. \& {Lin}, D.~N.~C. 2004, \apj, 610, 477

\bibitem[{{Ogilvie} \& {Lin}(2007)}]{Ogilvie_and_Lin_2007}
---. 2007, \apj, 661, 1180

\bibitem[{{Peale} \& {Cassen}(1978)}]{Peale_and_Cassen_1978}
{Peale}, S.~J. \& {Cassen}, P. 1978, Icarus, 36, 245

\bibitem[{{Rasio} {et~al.}(1996){Rasio}, {Tout}, {Lubow}, \&
  {Livio}}]{rasio_1996}
{Rasio}, F.~A., {Tout}, C.~A., {Lubow}, S.~H., \& {Livio}, M. 1996, \apj, 470,
  1187

\bibitem[{{Saumon} {et~al.}(1995){Saumon}, {Chabrier}, \& {van
  Horn}}]{Saumon_et_al_1995}
{Saumon}, D., {Chabrier}, G., \& {van Horn}, H.~M. 1995, \apjs, 99, 713

\bibitem[{{Showman} {et~al.}(2008{\natexlab{a}}){Showman}, {Cooper}, {Fortney},
  \& {Marley}}]{showman_et_al2008}
{Showman}, A.~P., {Cooper}, C.~S., {Fortney}, J.~J., \& {Marley}, M.~S.
  2008{\natexlab{a}}, \apj, 682, 559

\bibitem[{{Showman} {et~al.}(2008{\natexlab{b}}){Showman}, {Fortney}, {Lian},
  {Marley}, {Freedman}, {Knutson}, \& {Charbonneau}}]{showman_et_al2008b}
{Showman}, A.~P., {Fortney}, J.~J., {Lian}, Y., {Marley}, M.~S., {Freedman},
  R.~S., {Knutson}, H.~A., \& {Charbonneau}, D. 2008{\natexlab{b}}, submitted
  to \apj~(arXiv:0809.2089)

\bibitem[{{Showman} \& {Guillot}(2002)}]{showman+guillot2002}
{Showman}, A.~P. \& {Guillot}, T. 2002, \aap, 385, 166

\bibitem[{{Showman} {et~al.}(2008{\natexlab{c}}){Showman}, {Menou}, \&
  {Cho}}]{showman_et_al_2007}
{Showman}, A.~P., {Menou}, K., \& {Cho}, J.~Y.-K. 2008{\natexlab{c}}, in
  Astronomical Society of the Pacific Conference Series, Vol. 398, Astronomical
  Society of the Pacific Conference Series, ed. D.~{Fischer}, F.~A. {Rasio},
  S.~E. {Thorsett}, \& A.~{Wolszczan}, 419--+

\bibitem[{{Sozzetti} {et~al.}(2009){Sozzetti}, {Torres}, {Charbonneau}, {Winn},
  {Korzennik}, {Holman}, {Latham}, {Laird}, {Fernandez}, {O'Donovan},
  {Mandushev}, {Dunham}, {Everett}, {Esquerdo}, {Rabus}, {Belmonte}, {Deeg},
  {Brown}, {Hidas}, \& {Baliber}}]{Sozzetti_et_al_2008}
{Sozzetti}, A., {Torres}, G., {Charbonneau}, D., {Winn}, J.~N., {Korzennik},
  S.~G., {Holman}, M.~J., {Latham}, D.~W., {Laird}, J.~B., {Fernandez}, J.,
  {O'Donovan}, F.~T., {Mandushev}, G., {Dunham}, E., {Everett}, M.~E.,
  {Esquerdo}, G.~A., {Rabus}, M., {Belmonte}, J.~A., {Deeg}, H.~J., {Brown},
  T.~N., {Hidas}, M.~G., \& {Baliber}, N. 2009, \apj, 691, 1145

\bibitem[{{Torres} {et~al.}(2008){Torres}, {Winn}, \&
  {Holman}}]{Torres_et_al_2008}
{Torres}, G., {Winn}, J.~N., \& {Holman}, M.~J. 2008, \apj, 677, 1324

\bibitem[{{Wilson} {et~al.}(2008){Wilson}, {Gillon}, {Hellier}, {Maxted},
  {Pepe}, {Queloz}, {Anderson}, {Collier Cameron}, {Smalley}, {Lister},
  {Bentley}, {Blecha}, {Christian}, {Enoch}, {Haswell}, {Hebb}, {Horne},
  {Irwin}, {Joshi}, {Kane}, {Marmier}, {Mayor}, {Parley}, {Pollacco}, {Pont},
  {Ryans}, {Segransan}, {Skillen}, {Street}, {Udry}, {West}, \&
  {Wheatley}}]{Wilson_et_al_2008}
{Wilson}, D.~M., {Gillon}, M., {Hellier}, C., {Maxted}, P.~F.~L., {Pepe}, F.,
  {Queloz}, D., {Anderson}, D.~R., {Collier Cameron}, A., {Smalley}, B.,
  {Lister}, T.~A., {Bentley}, S.~J., {Blecha}, A., {Christian}, D.~J., {Enoch},
  B., {Haswell}, C.~A., {Hebb}, L., {Horne}, K., {Irwin}, J., {Joshi}, Y.~C.,
  {Kane}, S.~R., {Marmier}, M., {Mayor}, M., {Parley}, N., {Pollacco}, D.,
  {Pont}, F., {Ryans}, R., {Segransan}, D., {Skillen}, I., {Street}, R.~A.,
  {Udry}, S., {West}, R.~G., \& {Wheatley}, P.~J. 2008, \apjl, 675, L113

\bibitem[{{Winn} \& {Holman}(2005)}]{winn+holman2005}
{Winn}, J.~N. \& {Holman}, M.~J. 2005, \apjl, 628, L159

\bibitem[{{Winn} {et~al.}(2007){Winn}, {Holman}, {Bakos}, {P{\'a}l}, {Johnson},
  {Williams}, {Shporer}, {Mazeh}, {Fernandez}, {Latham}, \&
  {Gillon}}]{Winn_et_al_2007}
{Winn}, J.~N., {Holman}, M.~J., {Bakos}, G.~{\'A}., {P{\'a}l}, A., {Johnson},
  J.~A., {Williams}, P.~K.~G., {Shporer}, A., {Mazeh}, T., {Fernandez}, J.,
  {Latham}, D.~W., \& {Gillon}, M. 2007, \aj, 134, 1707

\bibitem[{{Winn} {et~al.}(2009){Winn}, {Holman}, {Carter}, {Torres}, {Osip}, \&
  {Beatty}}]{Winn_et_al_2009_1}
{Winn}, J.~N., {Holman}, M.~J., {Carter}, J.~A., {Torres}, G., {Osip}, D.~J.,
  \& {Beatty}, T. 2009, \aj, 137, 3826

\bibitem[{{Winn} {et~al.}(2008){Winn}, {Holman}, {Torres}, {McCullough},
  {Johns-Krull}, {Latham}, {Shporer}, {Mazeh}, {Garcia-Melendo}, {Foote},
  {Esquerdo}, \& {Everett}}]{Winn_et_al_2008}
{Winn}, J.~N., {Holman}, M.~J., {Torres}, G., {McCullough}, P., {Johns-Krull},
  C., {Latham}, D.~W., {Shporer}, A., {Mazeh}, T., {Garcia-Melendo}, E.,
  {Foote}, C., {Esquerdo}, G., \& {Everett}, M. 2008, \apj, 683, 1076

\bibitem[{{Wu}(2003)}]{Wu_2003}
{Wu}, Y. 2003, in Astronomical Society of the Pacific Conference Series, Vol.
  294, Scientific Frontiers in Research on Extrasolar Planets, ed. D.~{Deming}
  \& S.~{Seager}, 213--216

\bibitem[{{Wu}(2005{\natexlab{a}})}]{Wu_2005_1}
{Wu}, Y. 2005{\natexlab{a}}, \apj, 635, 674

\bibitem[{{Wu}(2005{\natexlab{b}})}]{Wu_2005_2}
---. 2005{\natexlab{b}}, \apj, 635, 688

\bibitem[{{Wu} \& {Murray}(2003)}]{Wu_and_Murray_2003}
{Wu}, Y. \& {Murray}, N. 2003, \apj, 589, 605

\bibitem[{{Wu} {et~al.}(2007){Wu}, {Murray}, \& {Ramsahai}}]{Wu_et_al_2007}
{Wu}, Y., {Murray}, N.~W., \& {Ramsahai}, J.~M. 2007, \apj, 670, 820

\bibitem[{{Yoder} \& {Peale}(1981)}]{Yoder_and_Peale_1981}
{Yoder}, C.~F. \& {Peale}, S.~J. 1981, Icarus, 47, 1

\end{thebibliography}

\clearpage
\begin{table}[ht]
\begin{center}
\caption[Values of $K_{zz}$.]{Observational Data of the HD~209458 System.} \label{tab:HD209458_data}
\vspace{0.4in}
\begin{tabular}{p{3cm}ccccc}
  \tableline 
  \tableline
  Planet             &             a                    &     e      &  period     &       $\rm M_{p}$           &     $\rm R_{p}$    \\
                     &            (AU)                  &   (95.4\% confidence)          &  (days)     &       ($\rm M_{J}$)         &    ($\rm R_{J}$)   \\
  \tableline \\[0.0cm]
   HD~209458b        &  $0.04707^{+0.00046}_{-0.00047}$ & $< 0.028 $ & $3.524746$  &  $0.685^{+0.015}_{-0.014}$  &  $1.320^{+0.024}_{-0.025}$        \\[0.5cm]
  \tableline

\end{tabular}

\vspace{2cm}

\begin{tabular}{p{3cm}ccccc}
  \tableline
  \tableline
  Star               &        $\rm M_{\ast}$         &          $\rm R_{\ast}$         &      $\rm T_{\ast}$       &      $\rm \left[ Fe/H \right]_{\ast}$   &        age          \\
                     &        $\rm M_{\sun}$         &          $\rm R_{\sun}$         &          (K)             &                 (dex)                   &       (Gyr)         \\
  \tableline \\[0.0cm]
   HD~209458         &  $1.101^{+0.066}_{-0.062}$    &    $1.125^{+0.020}_{-0.023}$    &    $6065^{+50}_{-50}$    &         $0.00^{+0.05}_{-0.05}$          &    $3.1^{+0.8}_{-0.7}$   \\[0.5cm]
  \tableline

\end{tabular}
\tablecomments{Data are from \citet{knutson_et_al2007a}, \citet{Torres_et_al_2008}, and \citet{Madhusudhan_and_Winn_2008}.}
\end{center}
\end{table}

\clearpage
\begin{landscape}
\begin{figure}[ht]

\centerline{
\includegraphics[width=11.0cm,angle=0,clip=true]{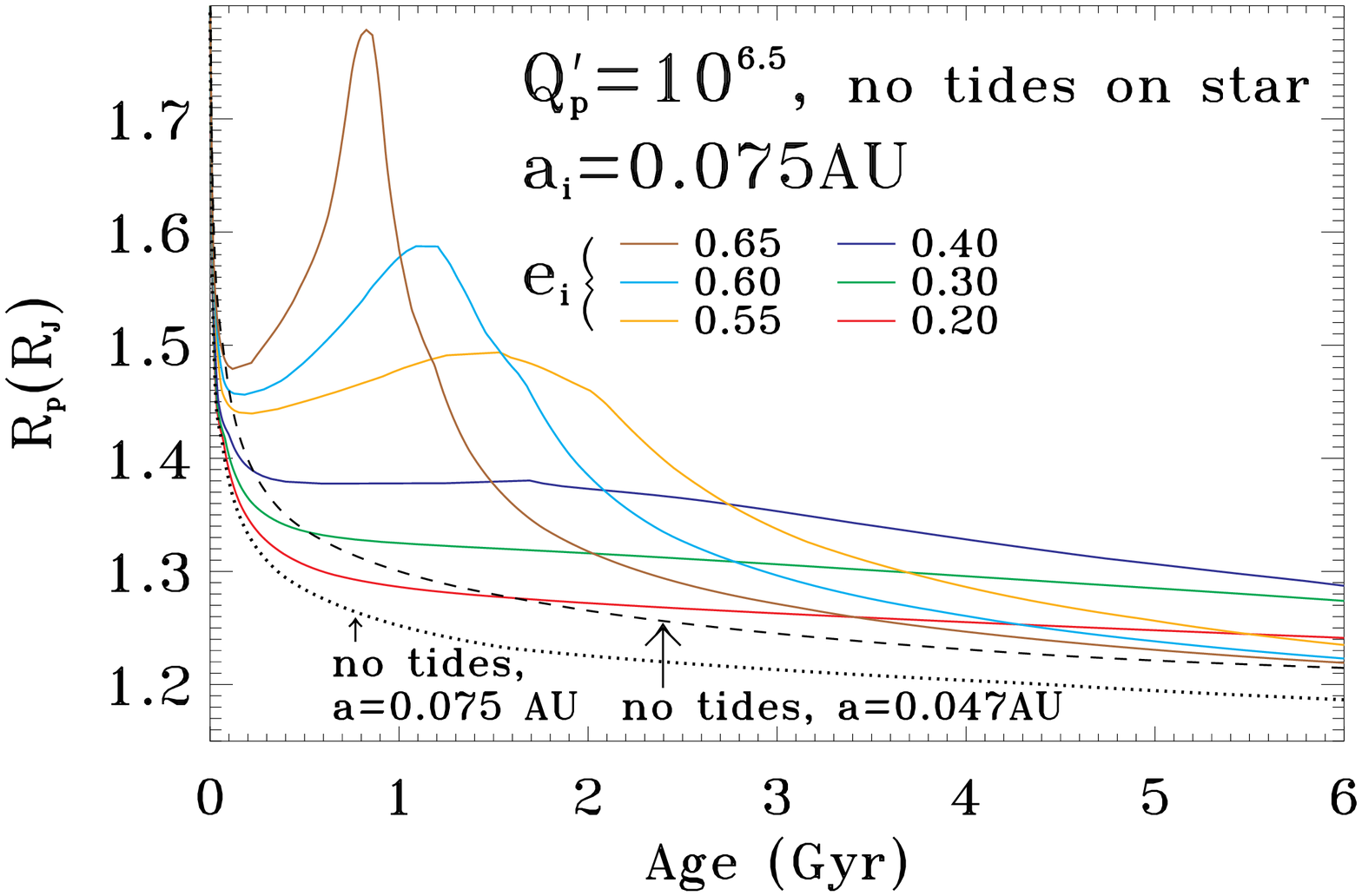}
\includegraphics[width=11.0cm,angle=0,clip=true]{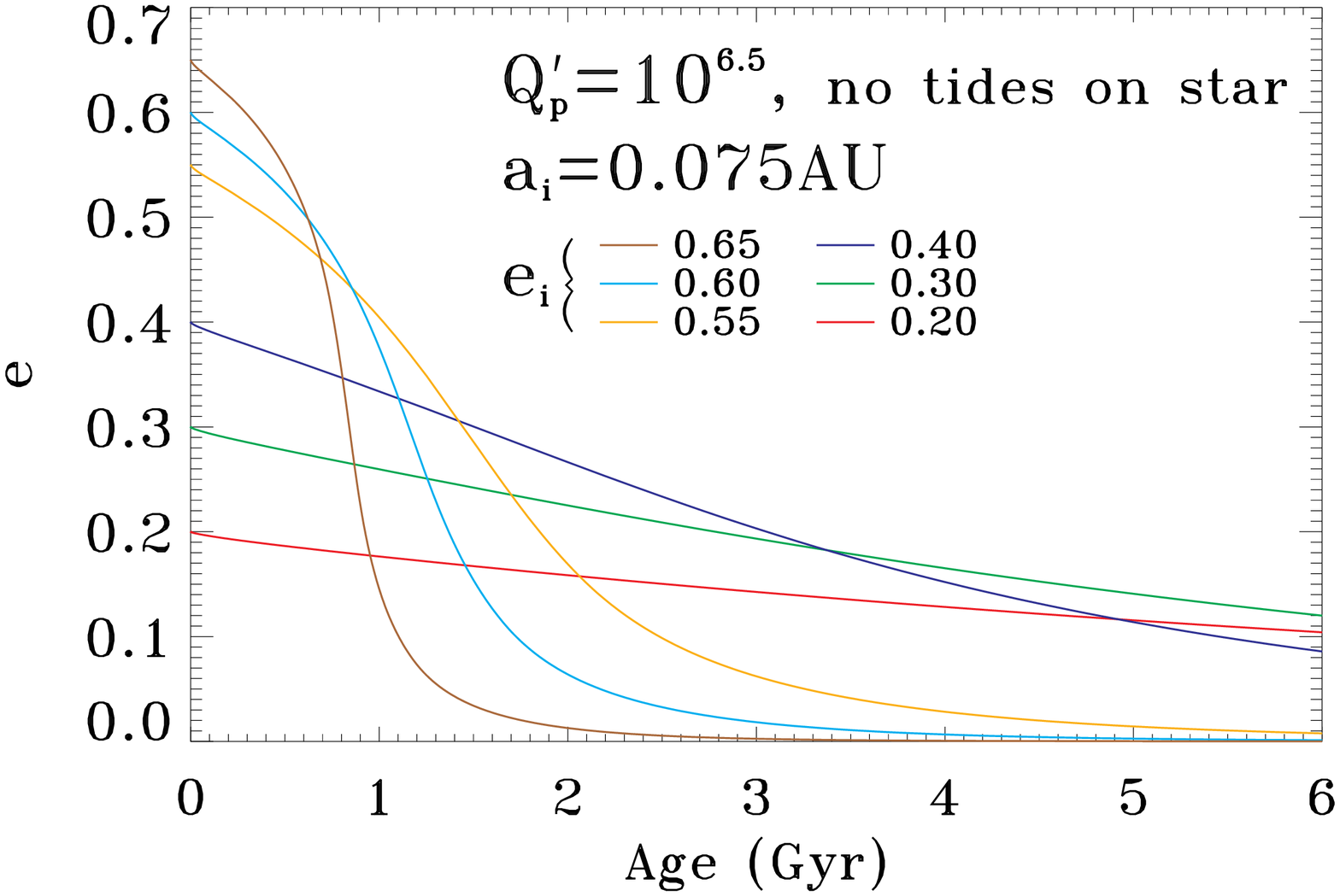}}
\centerline{
\includegraphics[width=11.0cm,angle=0,clip=true]{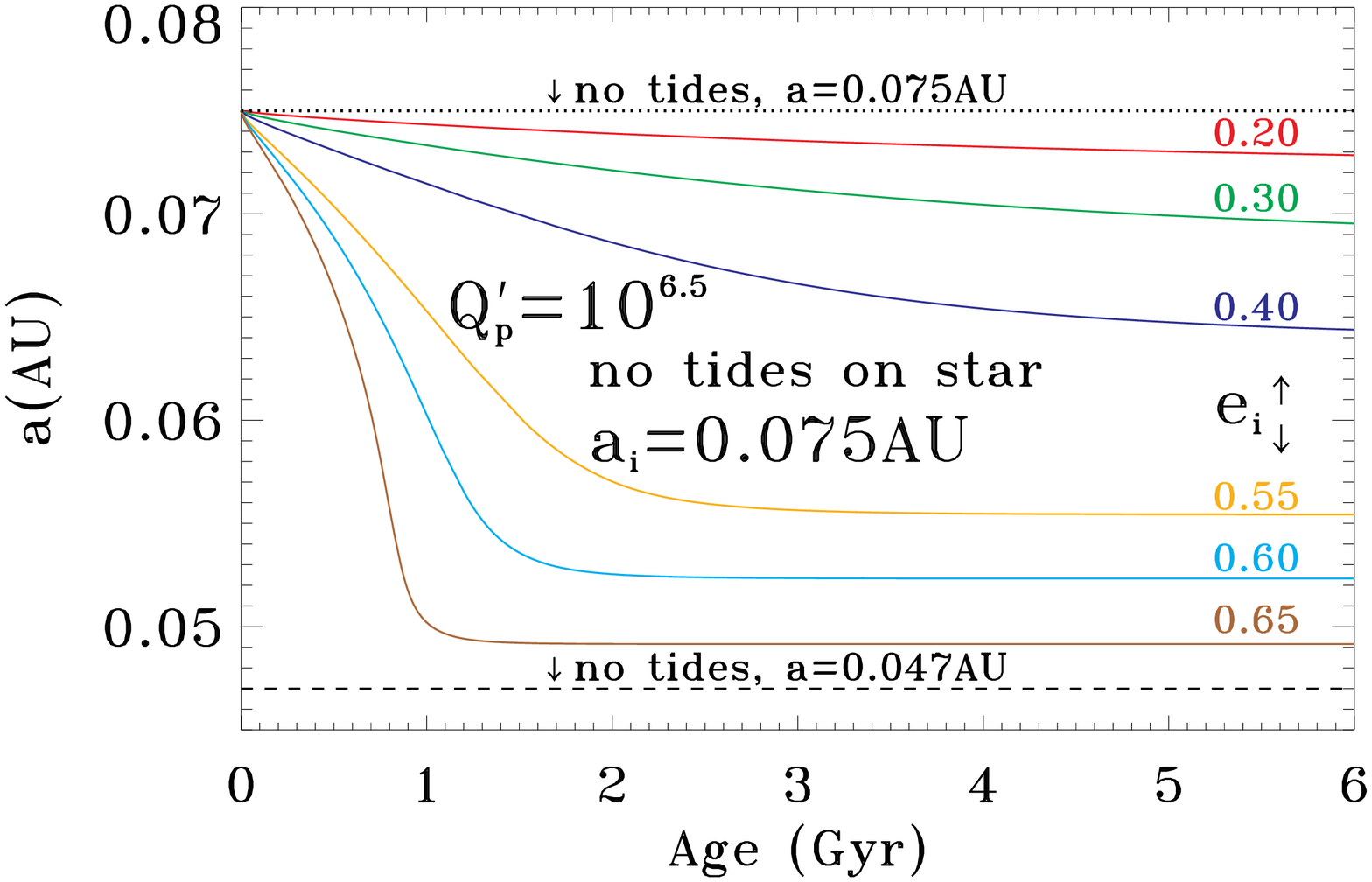}
\includegraphics[width=11.0cm,angle=0,clip=true]{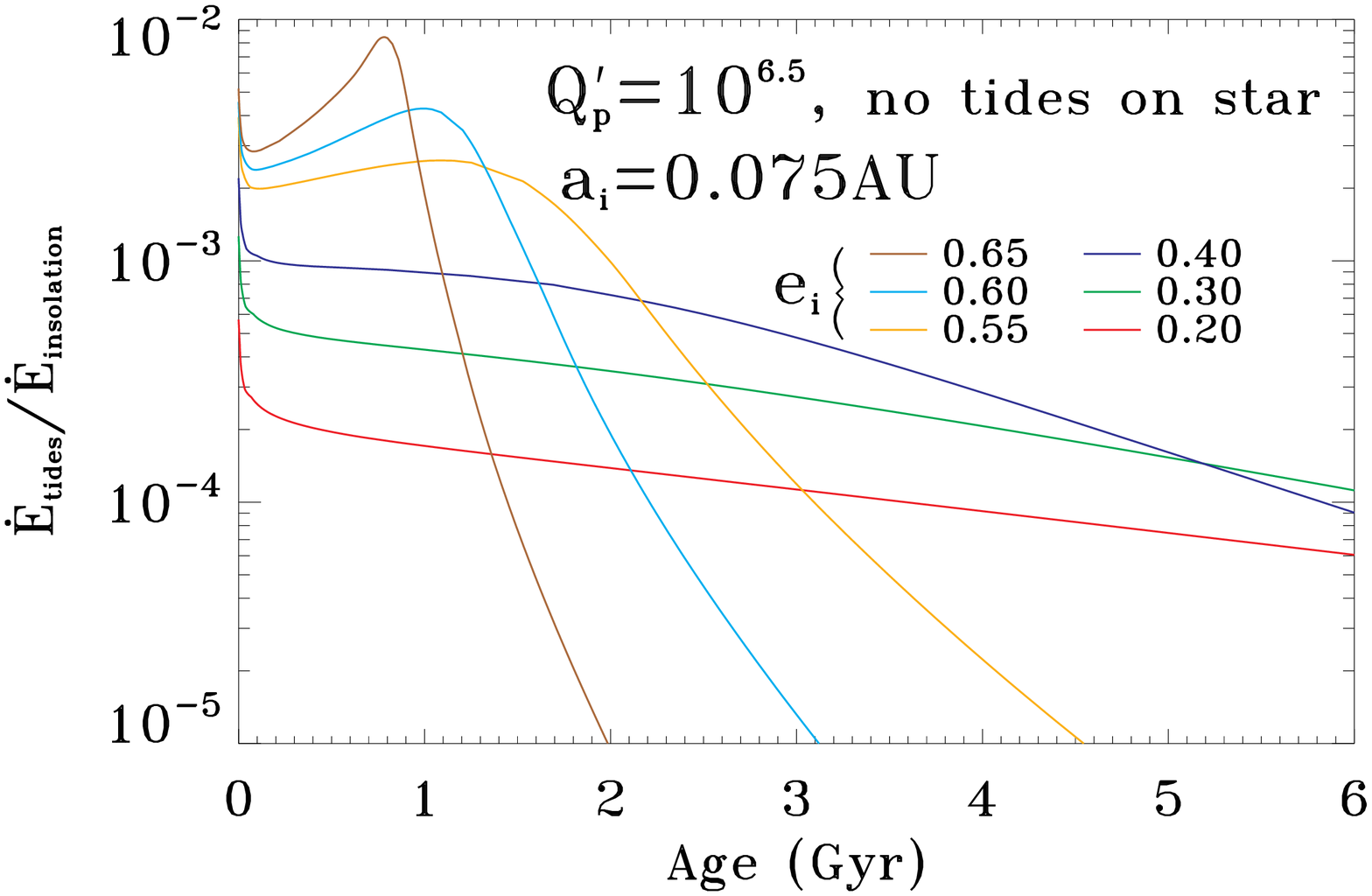}}

\caption{\footnotesize The Baseline scenario.
          Shown are the evolution of the planet radius $R_{p}~(R_{J})$ (top left), eccentricity $e$ (top right), semi-major axis $a$~(AU) (bottom left), 
          and power ratio $\dot{E}_{\rm tide}/ \dot{E}_{\rm insolation}$ (bottom right) versus age (in Gyr) for a representative planet (taken to be HD~209458b)
          at solar atmospheric opacity. $Q'_{p}$ is set equal to $10^{6.5}$
          and $a_{i}$ is set equal to 0.075 AU. We neglect tides raised on the star ($Q'_{\ast} \rightarrow \infty$). 
          $e_{i}$ assumes values of 0.2, 0.3, 0.4, 0.55, 0.60, and 0.65 (different colors).
	  Also shown in the left panels are radius evolution curves with no tides $-$ $a=a_{i}$ (dotted) and 
          $a=a_{\rm measured}=0.047$~AU (dashed). For $e_{i}=(0.65,0.60,0.55)$, the evolution undergoes
	  a transient phase of radius inflation. This phase does not appear for lower values of $e_{i}$ (0.40,0.30,0.20), 
          but the radius increase effect at later times is still in evidence.  The orbits eventually achieve circular 
          equilibrium states $-$ the higher the $e_{i}$, the closer to the star. See the discussion in \S\ref{subsec:typical_case}.
        }
\label{fig:ms_evolution_fig1}
\end{figure}

\clearpage
\begin{figure}[ht]
\centerline{
\includegraphics[width=11.0cm,angle=0,clip=true]{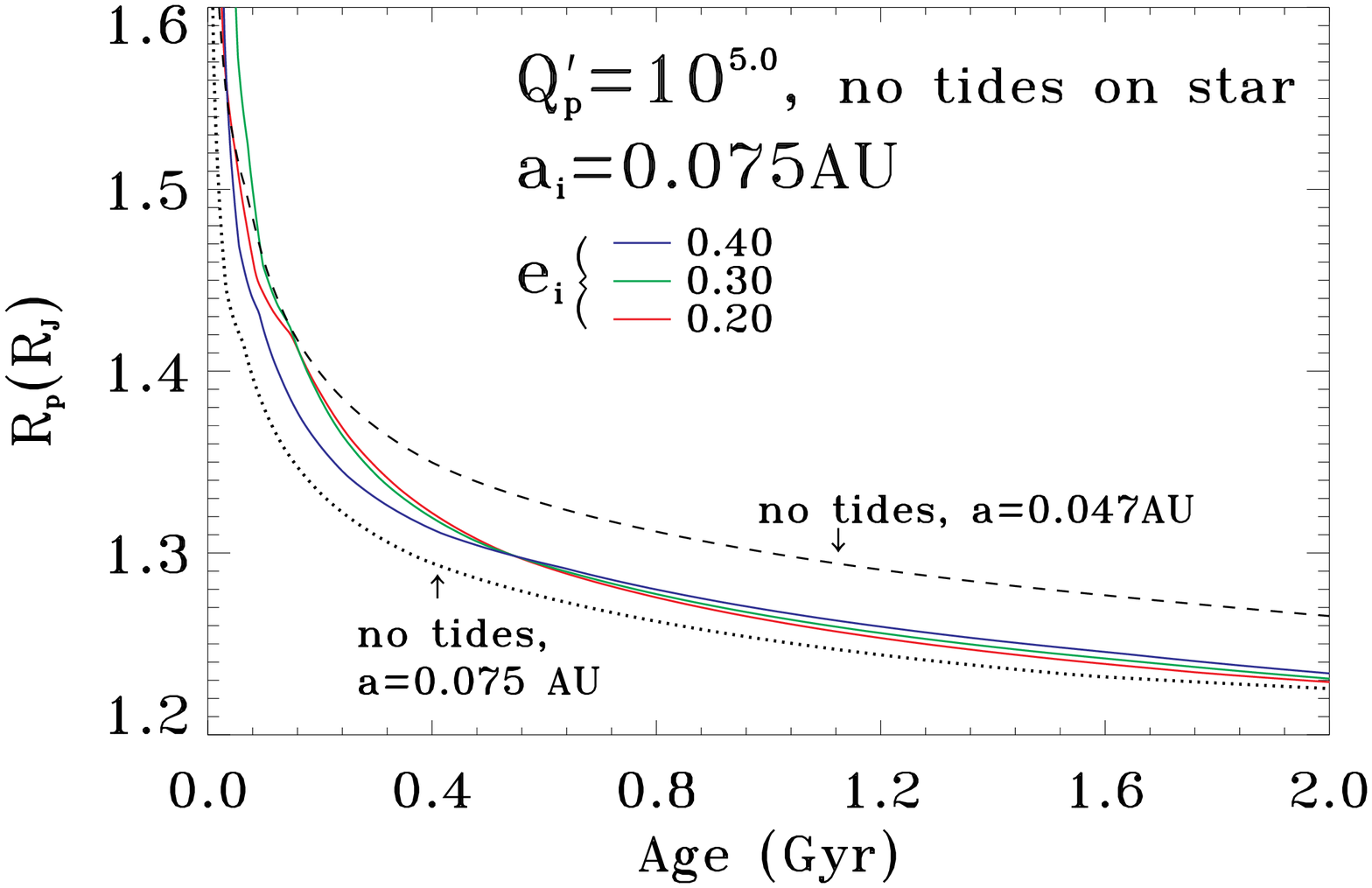}
\includegraphics[width=11.0cm,angle=0,clip=true]{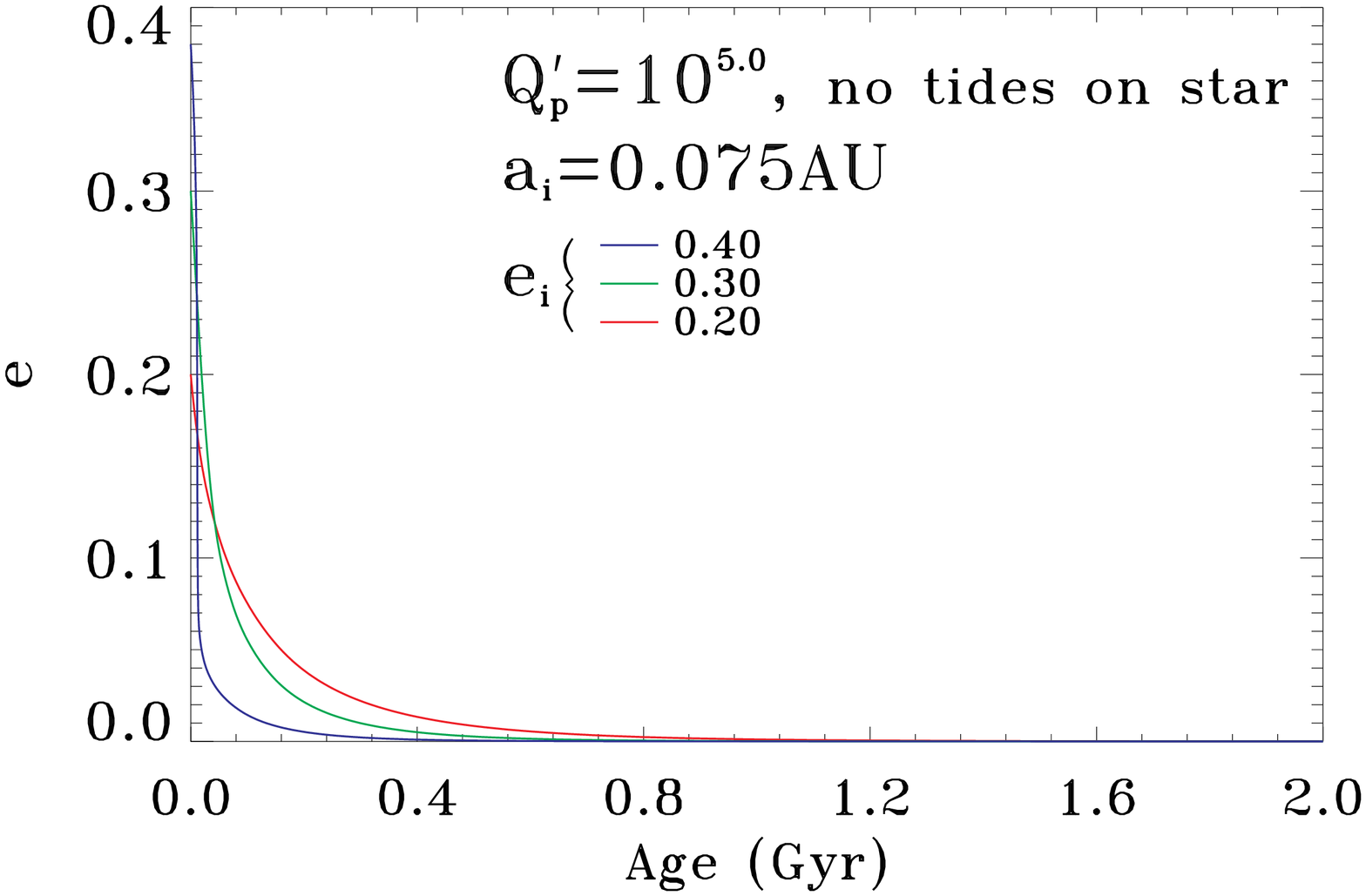}}
\centerline{
\includegraphics[width=11.0cm,angle=0,clip=true]{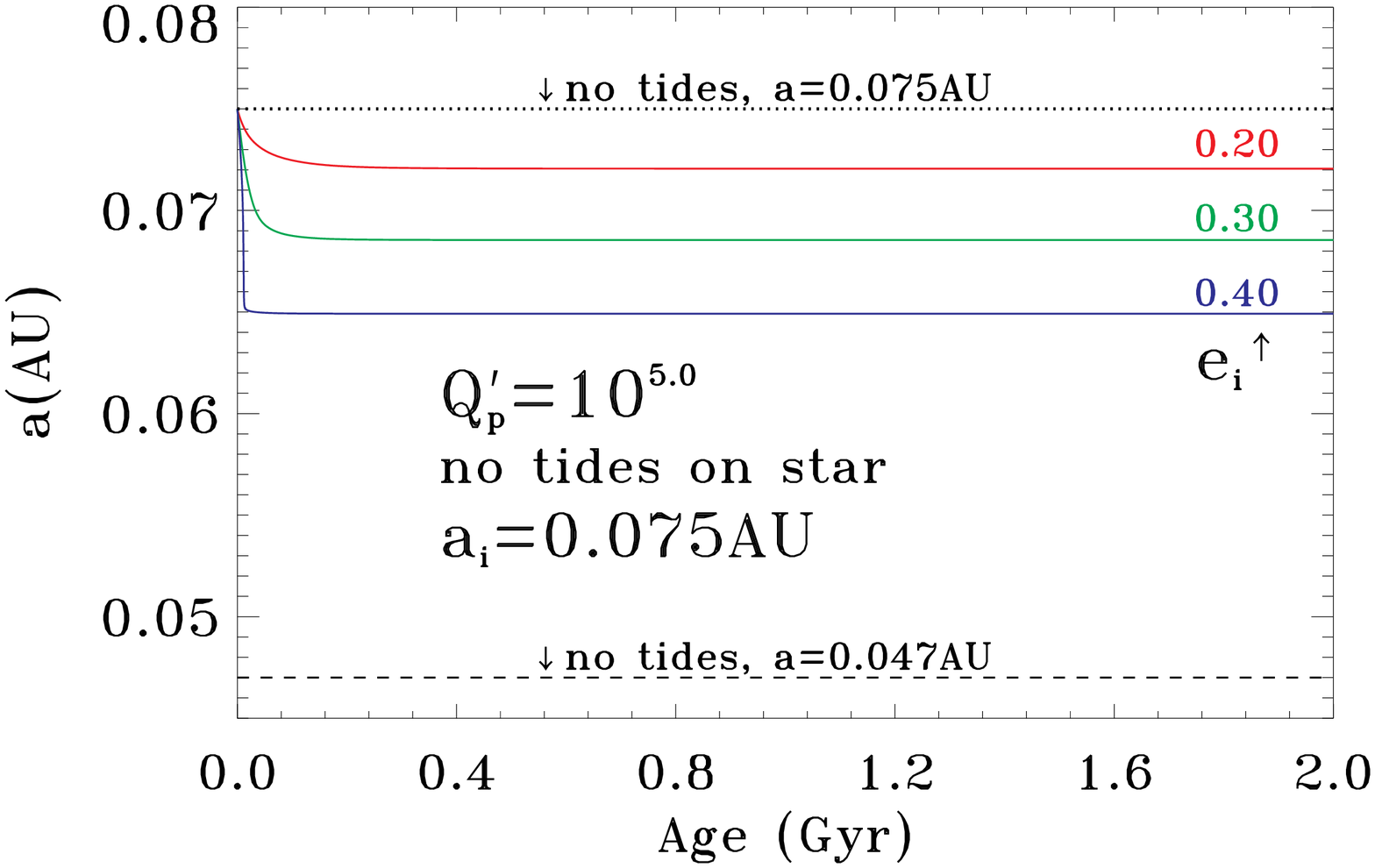}
\includegraphics[width=11.0cm,angle=0,clip=true]{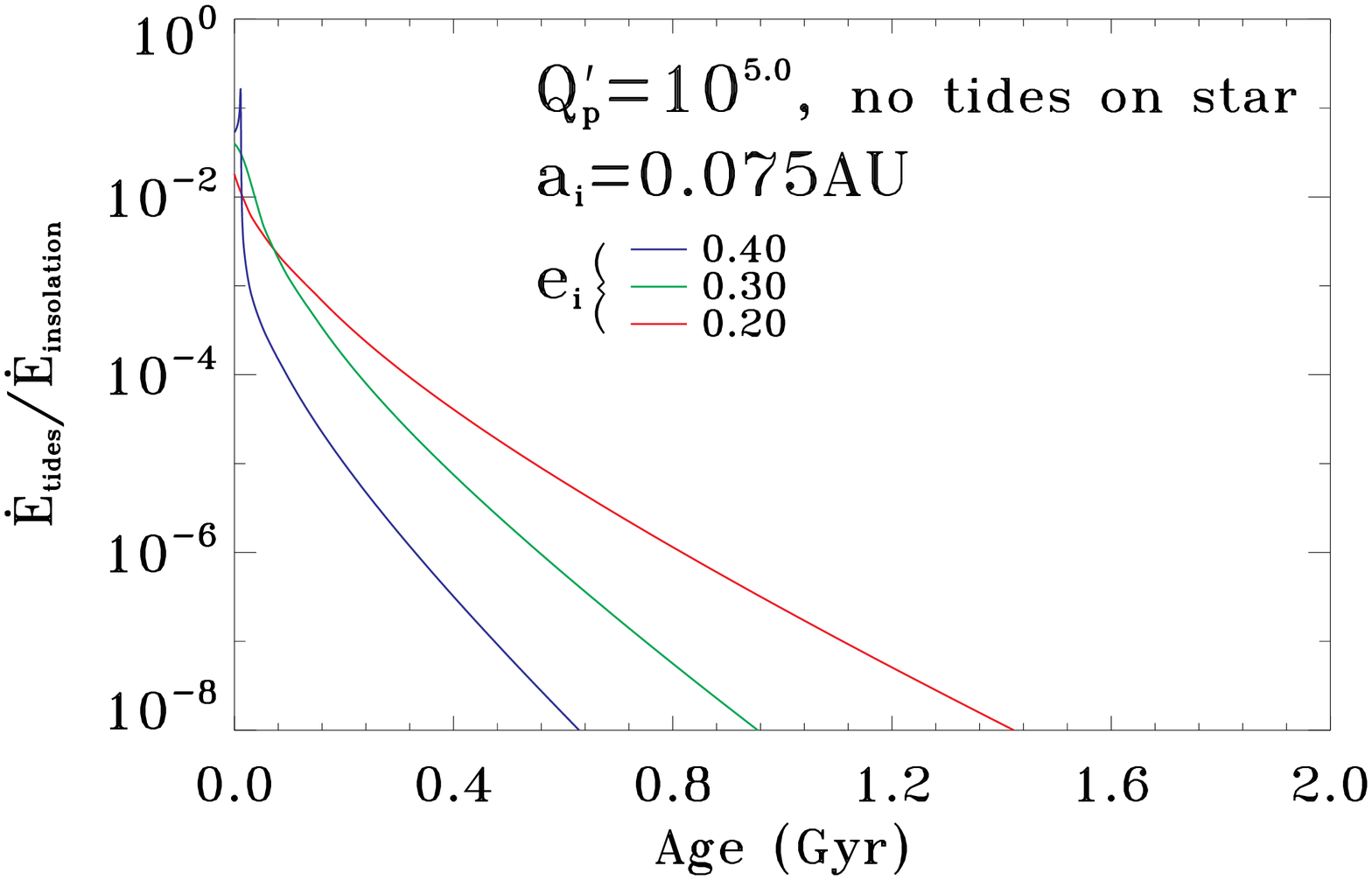}}

\caption{\footnotesize The same as in Fig. \ref{fig:ms_evolution_fig1}, but for the case of very strong planetary tidal effects (at a low value of $Q'_{p}=10^5$)
          and for the first 2 Gyrs. Also shown on the two left panels are two reference radius evolution curves incorporating 
          no tides $-$ one for $a=a_{i}$ (dotted) and one for $a=a_{\rm measured}=0.047$~AU (dashed).
          Models for only a subset of values of $e_i$ (0.20, 0.30, 0.40) are depicted.  The tidal effects fade at very early stages, so that their effect 
          on radius evolution is quite small at the observed ages of transiting planets, a few Gyr. The orbits circularize 
	  in less than $\sim$0.8 Gyr. See the text in \S\ref{subsec:strong_tides} for a discussion.
        }
\label{fig:ms_evolution_fig2}
\end{figure}

\clearpage
\begin{figure}[ht]
\centerline{
\includegraphics[width=11.0cm,angle=0,clip=true]{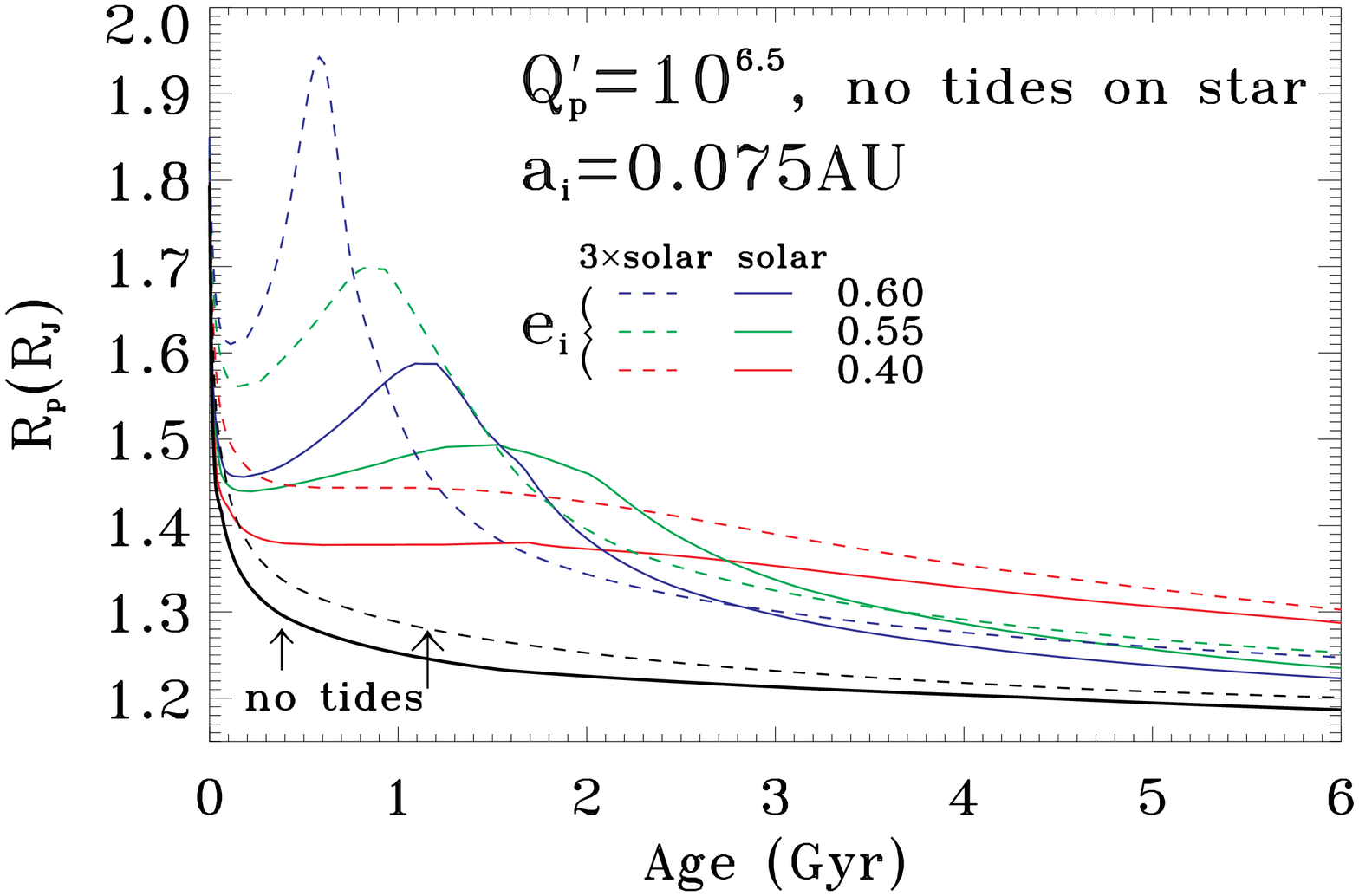}
\includegraphics[width=11.0cm,angle=0,clip=true]{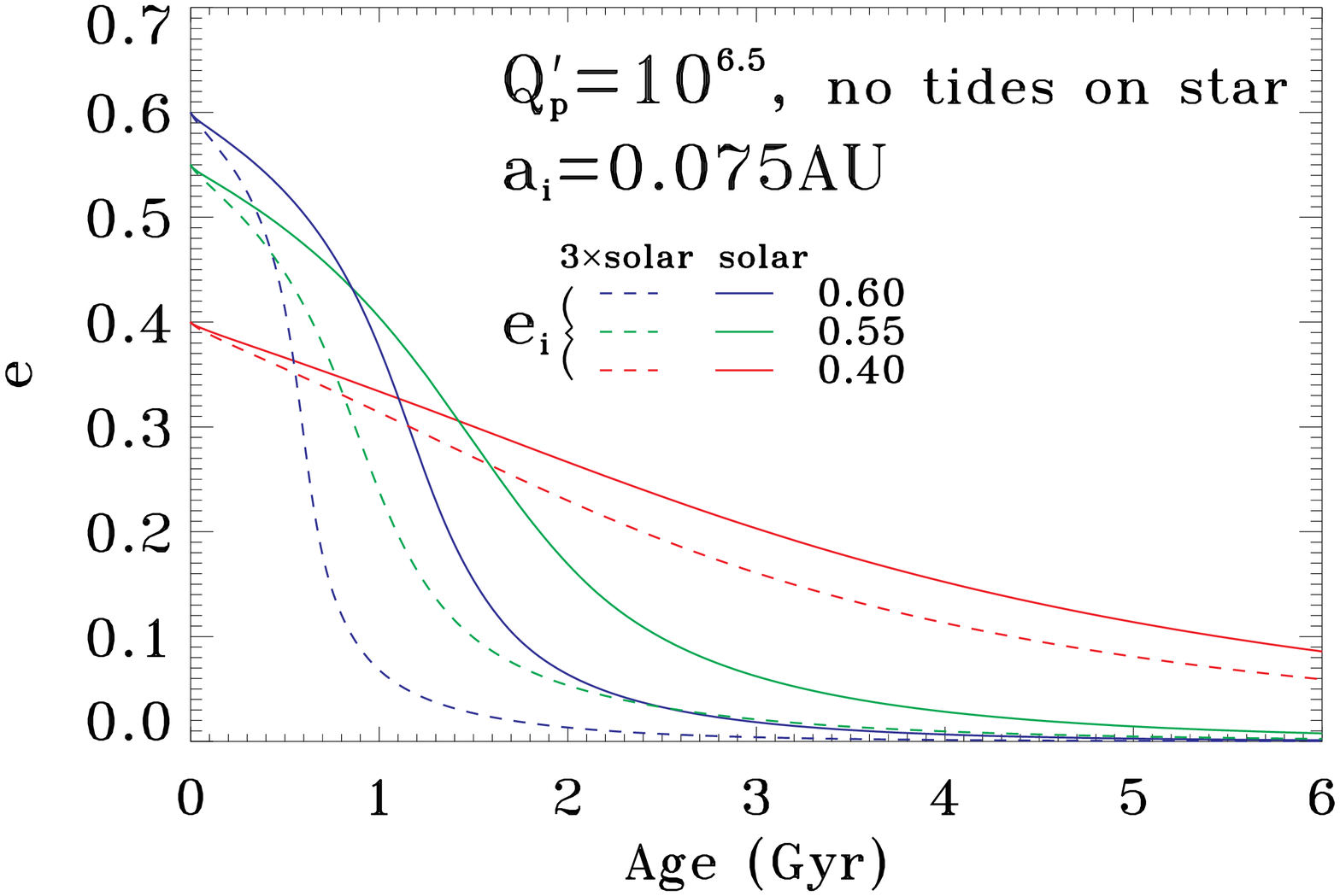}}
\centerline{
\includegraphics[width=11.0cm,angle=0,clip=true]{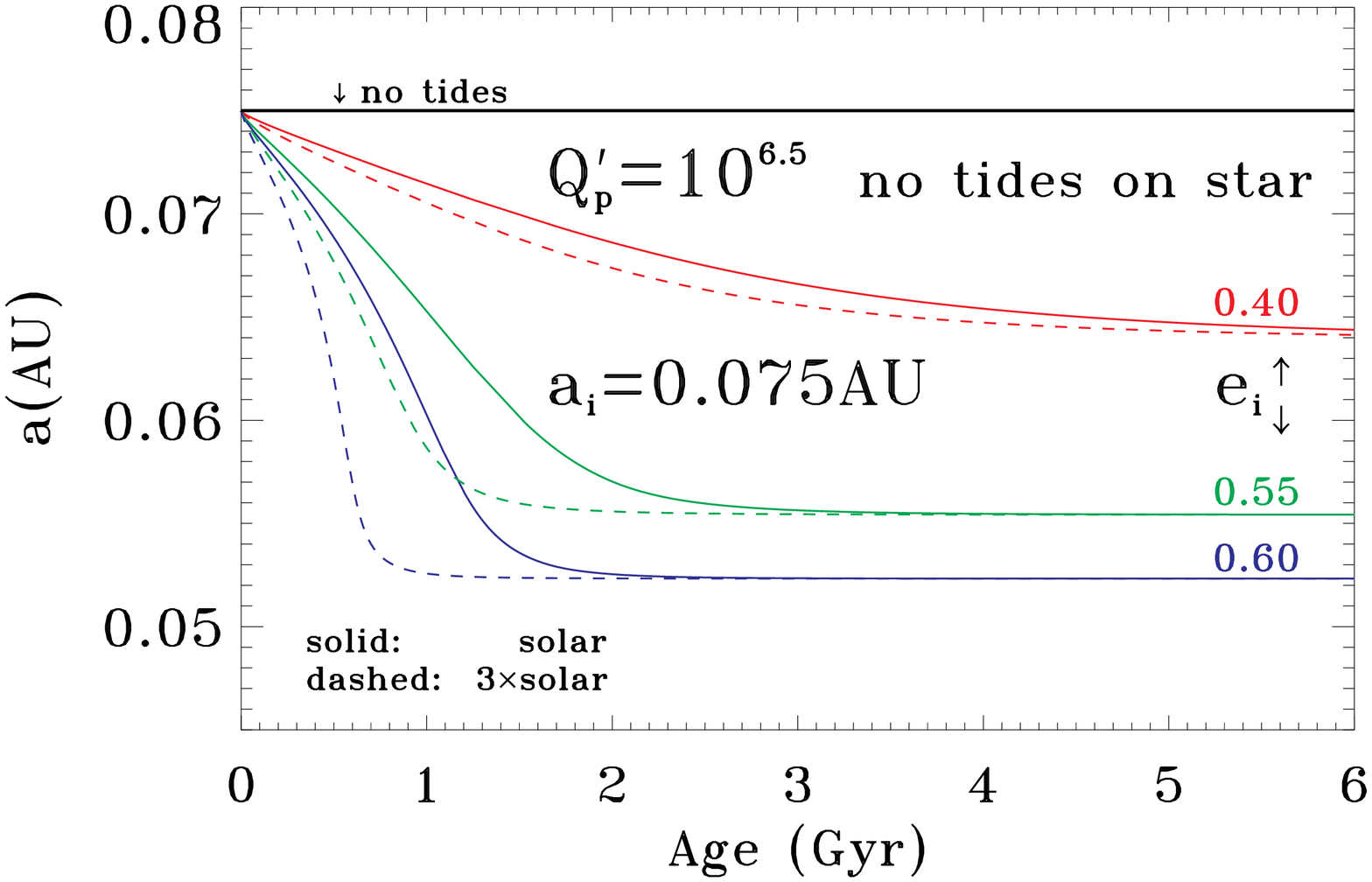}
\includegraphics[width=11.0cm,angle=0,clip=true]{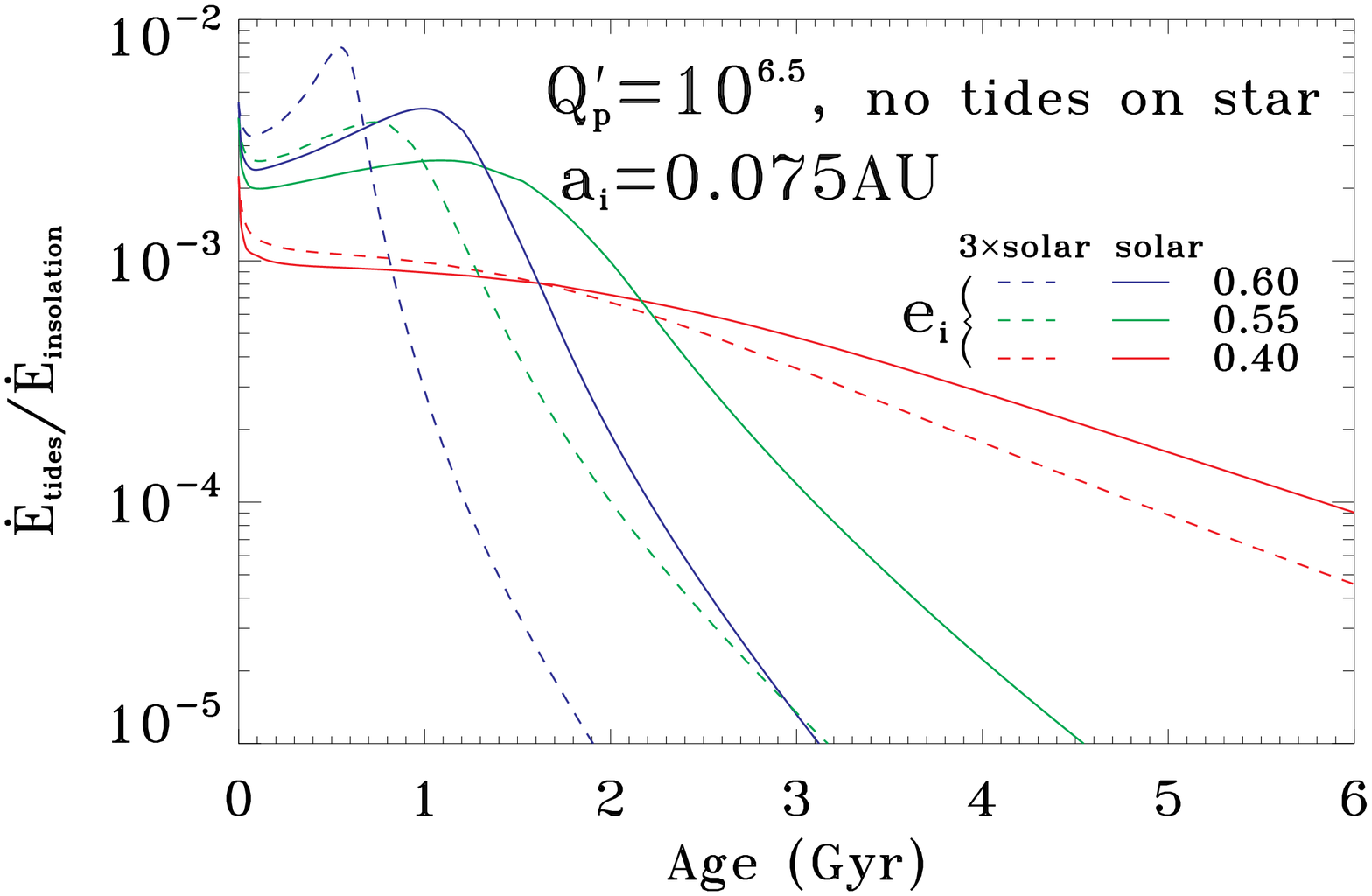}}
\caption{\footnotesize Same as in Fig. \ref{fig:ms_evolution_fig1}, but depicting the effect of atmospheric opacity. 
          Models for opacities with equilibrium chemical abundances at solar (solid) and 3$\times$solar (dashed) metallicity 
          are compared. The initial eccentricities $e_i$ are 0.40, 0.55, and 0.60 and they are distinguished by different colors.
          The planetary tidal dissipation factor ($Q'_{p}$) is set equal to 10$^{6.5}$ and the
	  tides raised on the star are neglected ($Q'_{\ast} \rightarrow \infty$). $a_i$ is set equal to 0.075 AU.
	  Also shown on the top left panel are two radius evolution curves with no tides $-$ for $a=a_{i}$ 
          at solar (solid) and 3$\times$solar (dashed) equilibrium opacities. The final orbital state for the 
          two opacities is the same, but that with higher opacity reaches it earlier. 
          See the text in \S\ref{subsec:atm_opacity} for a more complete discussion.
          }
\label{fig:ms_evolution_fig3}
\end{figure}

\clearpage
\begin{figure}[ht]
\centerline{
\includegraphics[width=11.0cm,angle=0,clip=true]{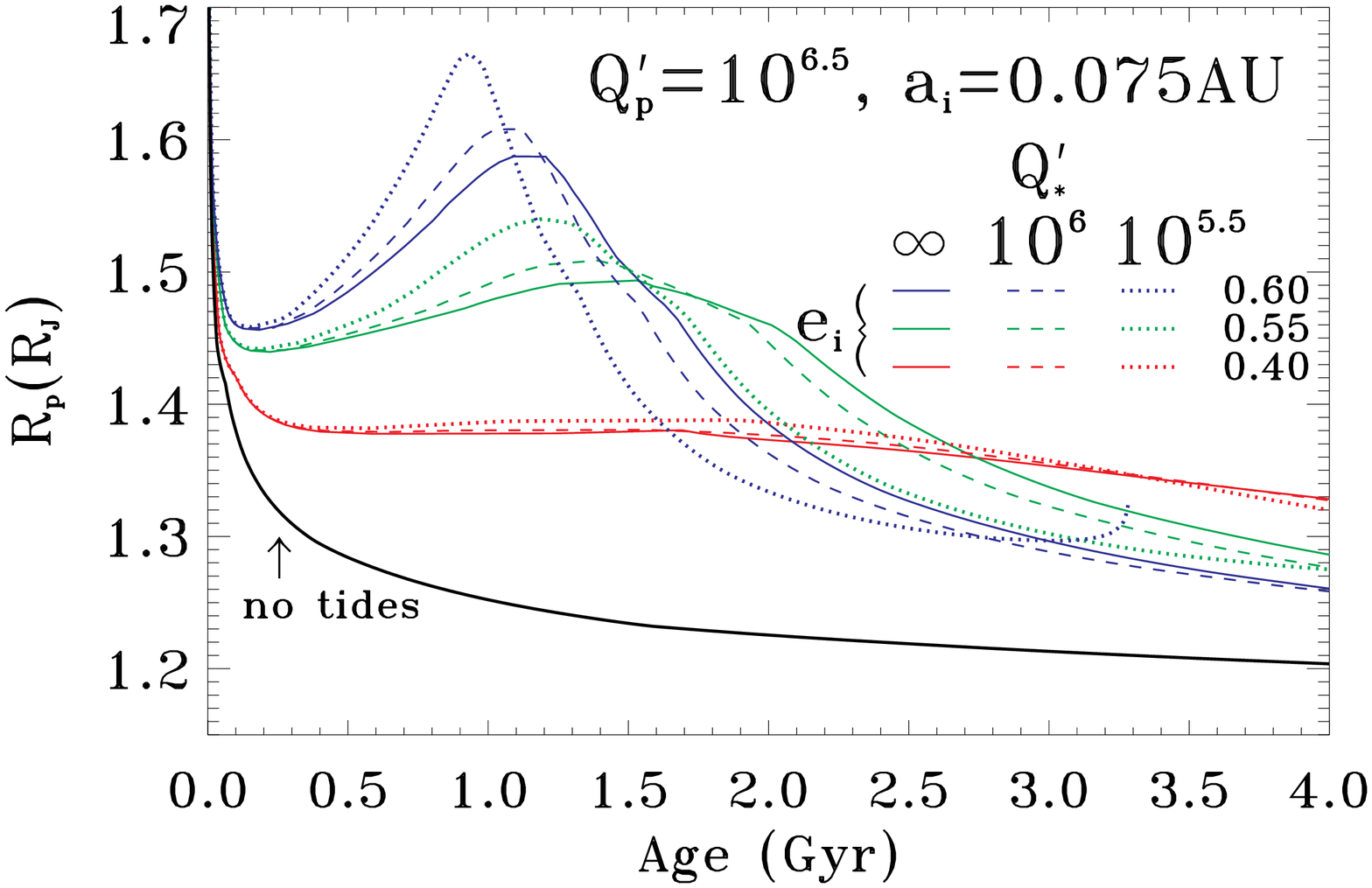}
\includegraphics[width=11.0cm,angle=0,clip=true]{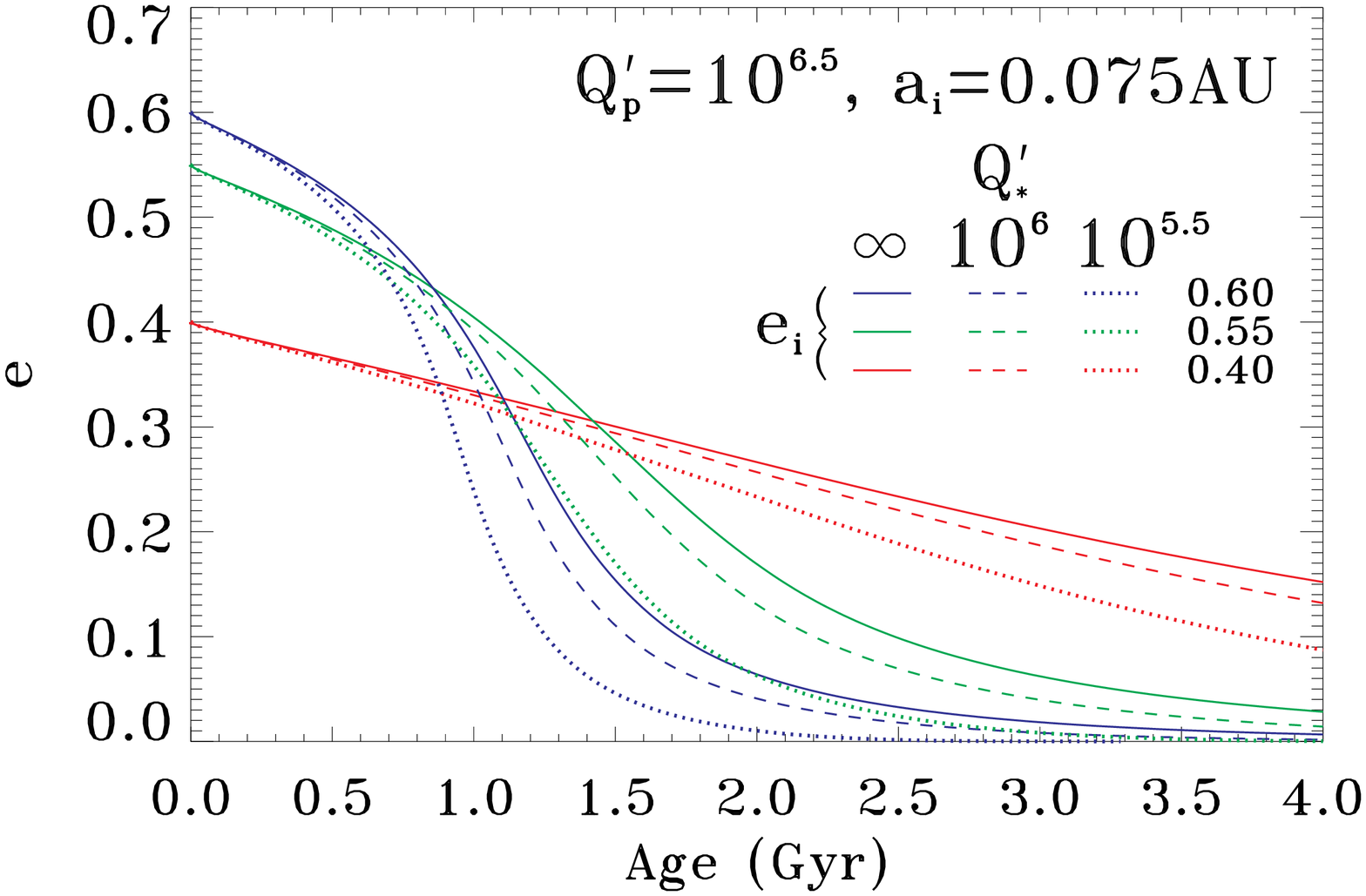}}
\centerline{
\includegraphics[width=11.0cm,angle=0,clip=true]{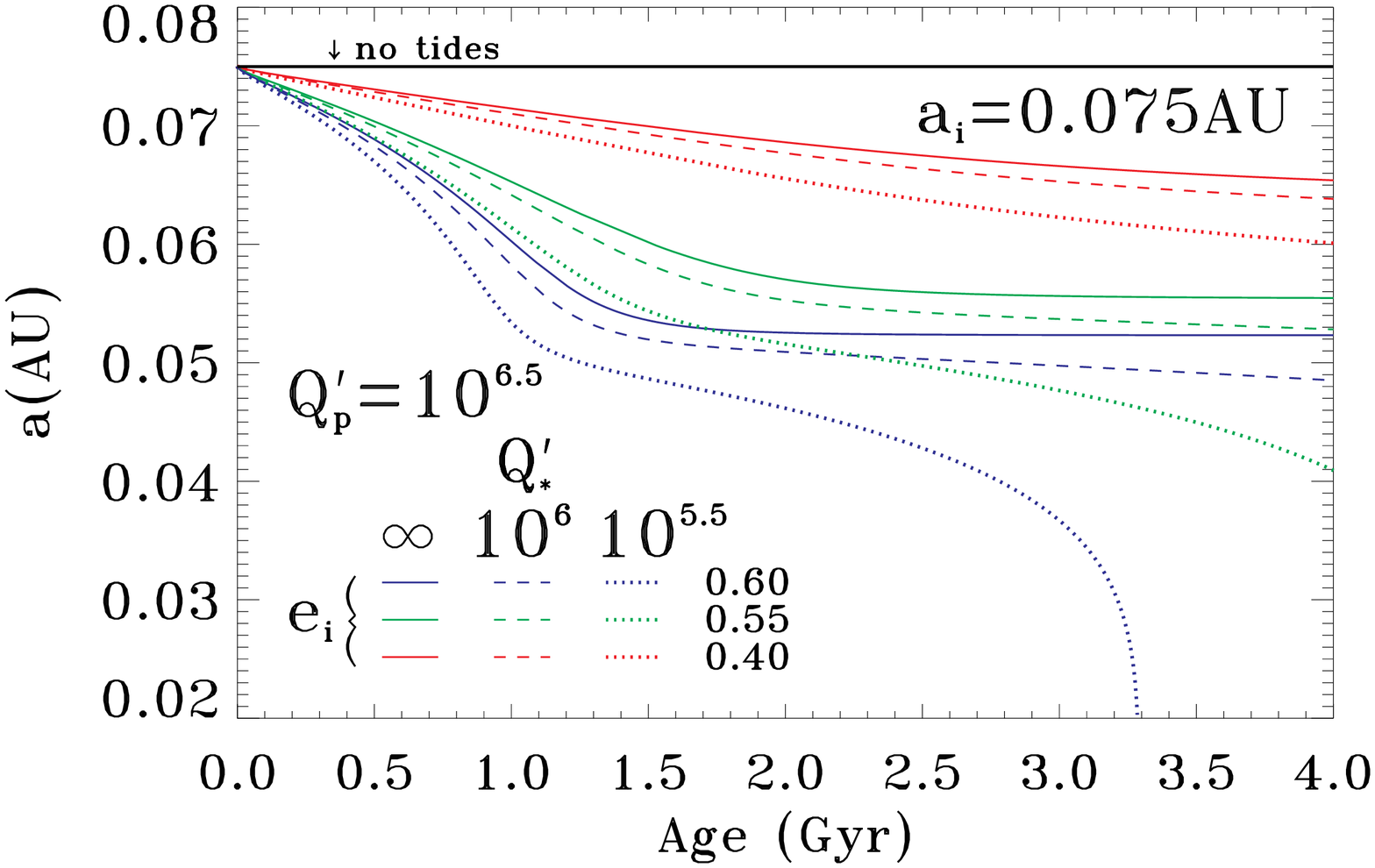}
\includegraphics[width=11.0cm,angle=0,clip=true]{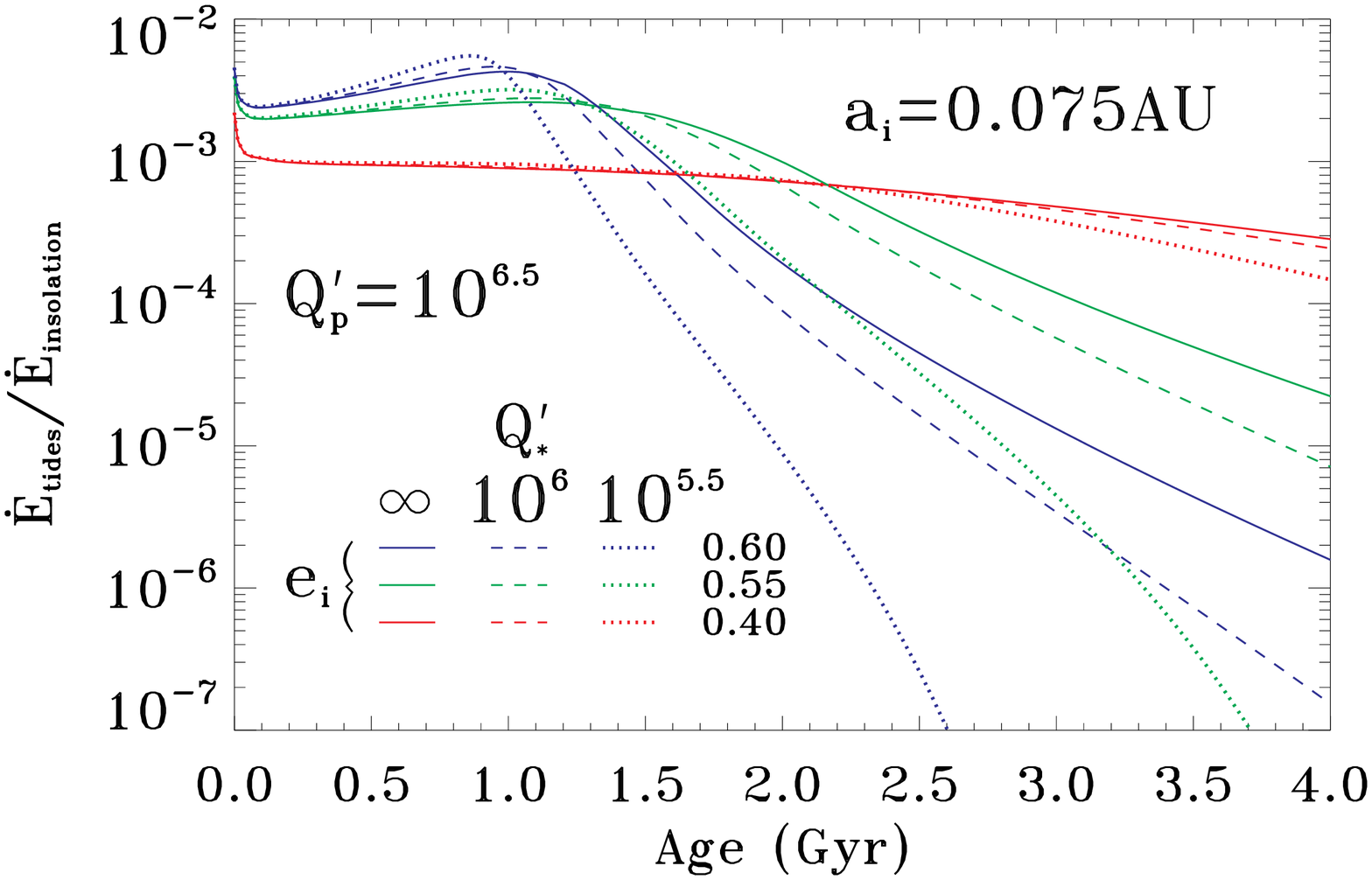}}
\caption{\footnotesize  Same as Fig. \ref{fig:ms_evolution_fig1}, but portraying the effect of tides raised on the star.
          Models for values of $Q'_{\ast}$ of $\infty$ (solid), 10$^{6.0}$ (dashed), and 10$^{5.5}$ (dotted) are compared. $e_i$ is set equal to 0.40, 0.50, 
          and 0.60 and these models are distinguished by different colors. $a_i = 0.075$~AU for all models and $Q'_{p}$ is the same for all models ($10^{6.5}$). 
          Also shown for reference in the two left panels is a model without tidal effects at all for $a=a_{i}$.
          Lower values of $Q'_{\ast}$ enhance and accelerate the transient phase of radius inflation. 
          The orbits generally evolve faster, eventually circularize,  but do not reach an equilibrium state.  The 
          evolution of $a$ for $Q'_{\ast} = 10^{5.5}$ and $e_i = 0.60$ (purple dotted curve) indicates that a 
          planet with such parameters eventually spirals into the star after $\sim$3.3 Gyrs (at an accelerating pace at later times). 
          See the text in \S\ref{subsec:tides_star} for a discussion.
        }
\label{fig:ms_evolution_fig4}
\end{figure}

\clearpage
\begin{figure}[ht]
\centerline{
\includegraphics[width=11.0cm,angle=0,clip=true]{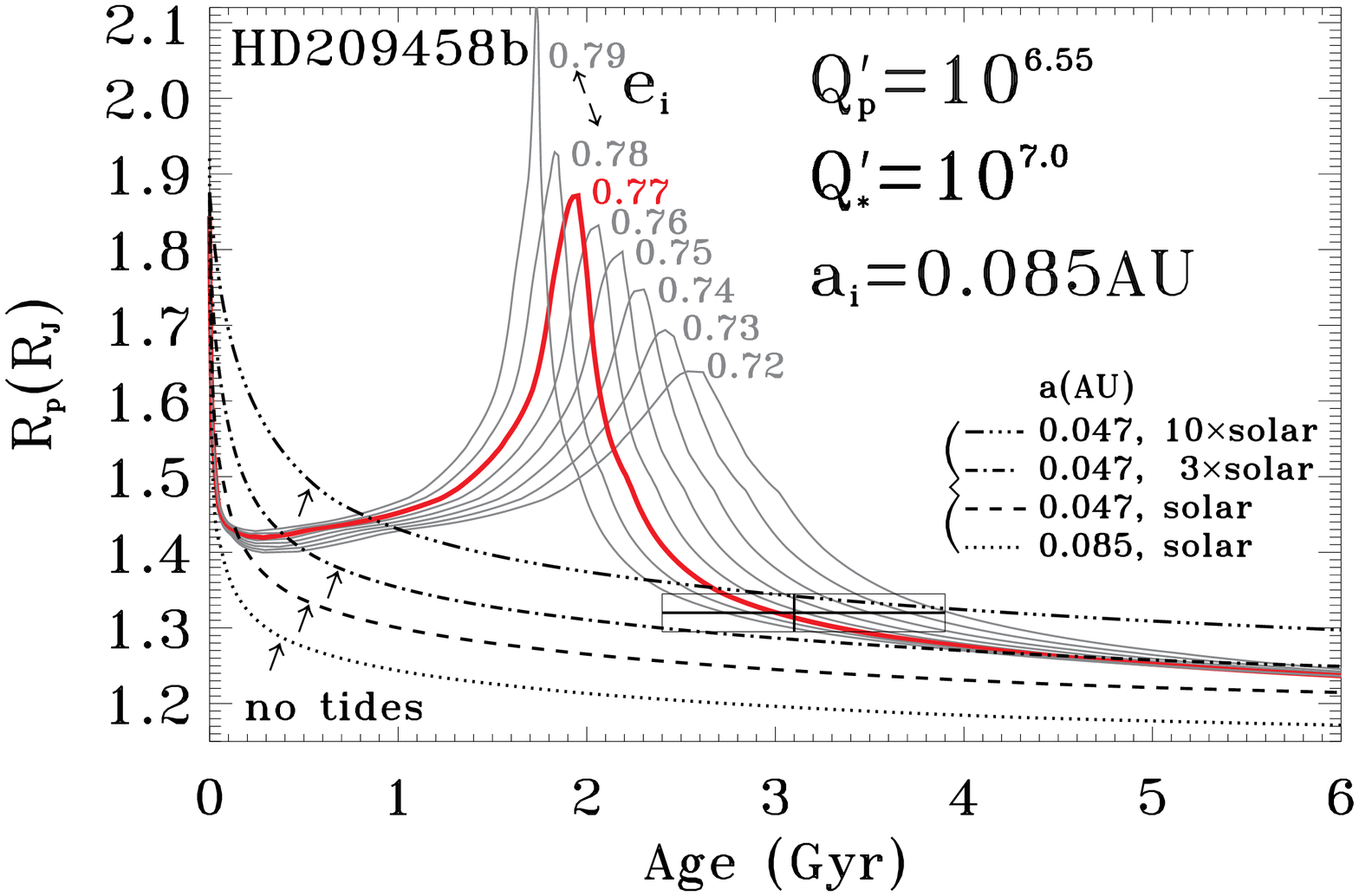}
\includegraphics[width=11.0cm,angle=0,clip=true]{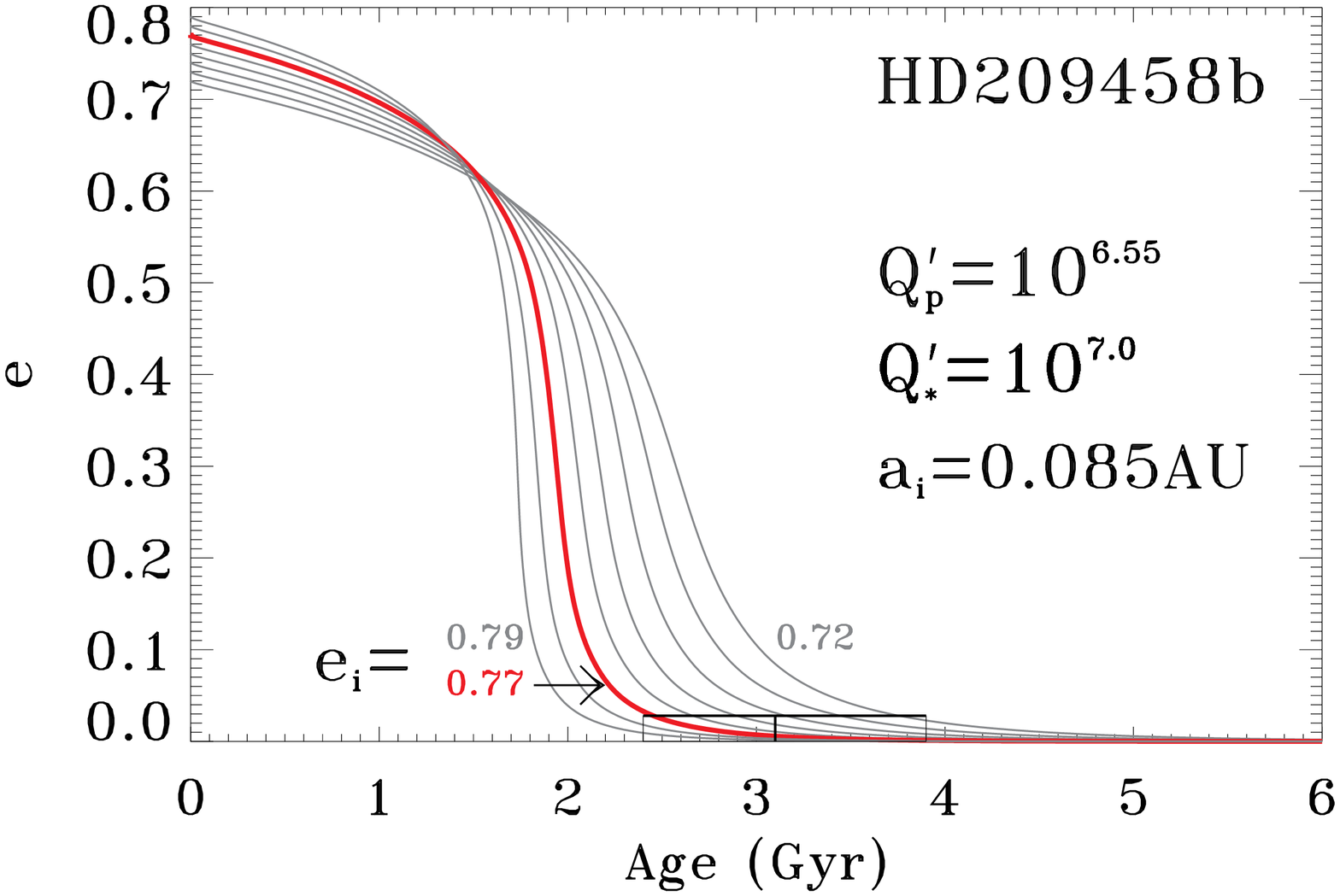}}
\centerline{
\includegraphics[width=11.0cm,angle=0,clip=true]{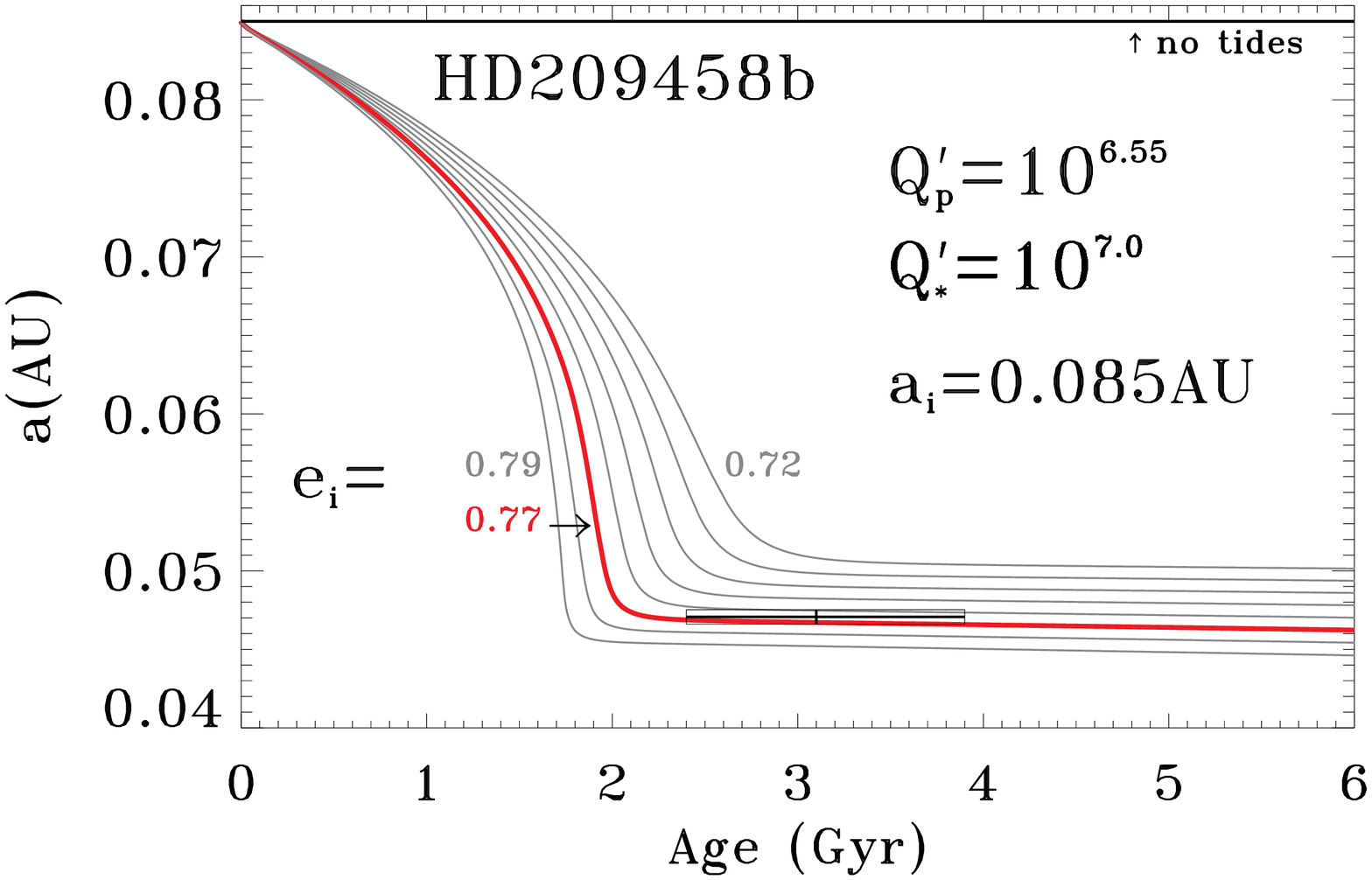}
\includegraphics[width=11.0cm,angle=0,clip=true]{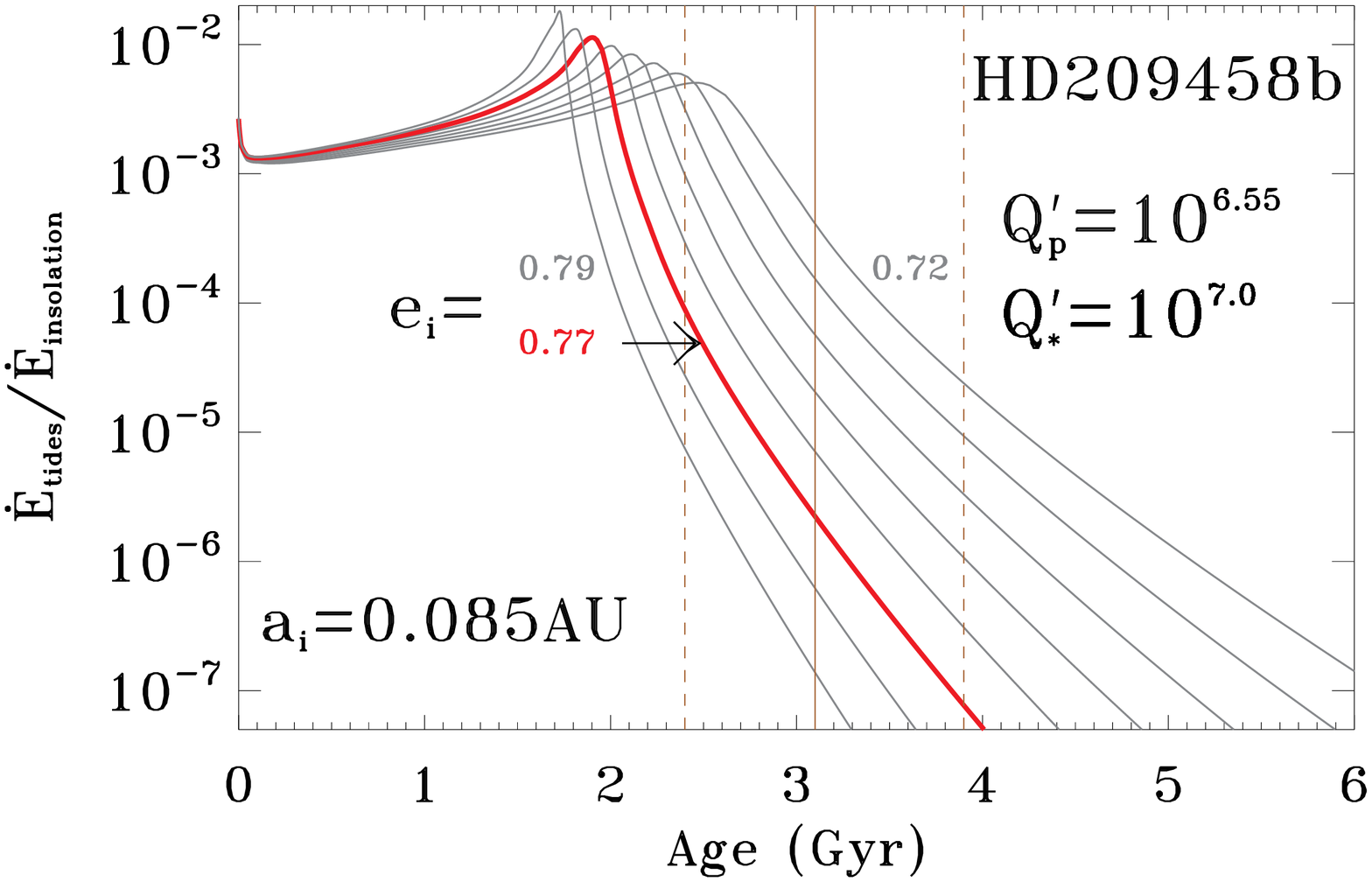}}
\caption{\footnotesize Example fit of the observed values of $R_p$, $e$, and $a$ for HD~209458b. 
          The measured $R_{p}$, $e$, and $a$ are shown with error boxes.
          The best fitting curve is for $e_{i}=0.77$, $Q'_{p} = 10^{6.55}$, 
          and $Q'_{\ast} = 10^{7.0}$.  This model is shown in red.  We have used 
          $a_i = 0.085$ AU.  To demonstrate the strong dependence on $e_i$, we include 
          models for $e_i$ between 0.72 and 0.79. Also shown in the top left panel are 
          four radius evolution curves (in non-solid black) that ignore the effects of tides.  They are 
          for $a=a_{i}$, but also for $a=a_{\rm measured}=0.047$~AU at solar, 
          3$\times$solar, and 10$\times$solar atmospheric opacities. The best 
          fitting curves without tidal effects are for $\gtrsim$3$\times$solar. 
          See the text in \S\ref{sec:HD209458b} for a discussion.
        }
\label{fig:ms_evolution_fig5}
\end{figure}

\clearpage

\end{landscape}

\end{document}